\documentclass[11pt]{article}
\usepackage[margin=1in]{geometry}
\usepackage{amsmath,amssymb,amsthm,mathtools}
\usepackage{booktabs}
\usepackage{tabularx,array}
\usepackage{pgfplots}
\usepgfplotslibrary{groupplots}
\pgfplotsset{compat=1.18}
\usepackage{xcolor}
\usepackage[round]{natbib}
\usepackage[hidelinks,hypertexnames=false]{hyperref}
\usepackage[capitalise]{cleveref}

\newtheorem{theorem}{Theorem}
\newtheorem{proposition}[theorem]{Proposition}
\newtheorem{lemma}[theorem]{Lemma}
\newtheorem{corollary}[theorem]{Corollary}
\theoremstyle{definition}
\newtheorem{definition}[theorem]{Definition}
\newtheorem{assumption}{Assumption}
\theoremstyle{remark}
\newtheorem{remark}[theorem]{Remark}
\crefname{assumption}{Assumption}{Assumptions}
\crefname{appendix}{Appendix}{Appendices}

\newcommand{\E}{\mathbb{E}}
\newcommand{\Var}{\operatorname{Var}}

\newcommand{\cM}{\mathcal{M}}
\newcommand{\expit}{\operatorname{expit}}
\newcommand{\logit}{\operatorname{logit}}
\newcommand{\hS}{\widehat S}
\newcommand{\htheta}{\hat\theta}
\newcommand{\hmtwo}{\hat m}
\newcommand{\indep}{\perp\!\!\!\perp}

\definecolor{cBlue}{HTML}{2A78D6}
\definecolor{cOrange}{HTML}{EB6834}
\definecolor{cAqua}{HTML}{1BAF7A}
\definecolor{cInk}{HTML}{52514E}

\title{Within-task hazards of AI agents:\\ what time-horizon curves, repeated runs\\ and step-level outcomes can identify}
\author{Ruiqi Zhao\\[2pt]
{\normalsize School of Data Science, The Chinese University of Hong Kong, Shenzhen}\\
{\normalsize\texttt{124090916@link.cuhk.edu.cn}}}
\date{November 2026}

\begin{document}
\maketitle

\begin{abstract}
An AI agent's time horizon is read off a curve: the probability that the agent completes a task, plotted against the time a human needs for it. Survival fits to this curve have been read as showing that agents' failure hazard declines within a task, which the curve cannot support: every log-logistic horizon curve with slope $\beta\le1$ is a mixture of constant-hazard curves, and in a proportional-hazards frailty model every within-task Weibull shape $k\ge\beta$ produces the same curve. With conditionally independent runs, repeated runs restore identification within that model: the probability that two runs both succeed is the single-run curve shifted by $1/k$ on the $\log_2$ time axis. Reading the result as a within-task hazard, however, needs a prefix assumption: a task's first steps must be as hard as when they are the whole task. End-of-task data cannot detect a violation of this assumption whenever they are compatible with it, whatever the estimator. Outcomes at a fixed checkpoint, on tasks with comparable prefixes, make it testable, and running the same instance at two lengths gives an exact paired test of the instance-level null. On a step-checkable task, in four preregistered experiments on independent instances with two open reasoning models from different families, asking for twice as many steps lowered average paired accuracy on the common first steps each time. Longer lists in the prompt were the largest part of the effect in point estimate on both models (144 more input tokens on the first; about half on the second). With the lists fixed, asking for more steps (which also drops an added instruction) had small point estimates on the first model and a positive effect on the second, even excluding unparsable runs. On the first model, a frailty correction from concurrent work gives shape intervals above within-run ones. We also reanalyse METR's run-level data under stated assumptions.
\end{abstract}

\section{Introduction}\label{sec:intro}

\citet{kwa2025measuring} summarise an agent's ability by its 50\% time horizon: fit a logistic curve of task success against $\log_2$ of the time humans need for the task, and report the human time at which predicted success is one half. Because the fitted curve declines with task length, it invites a survival reading. \citet{ord2025halflife} observed that a constant hazard per human-minute explains the data of \citet{kwa2025measuring} well, which gives each agent a half-life. Later analyses fitted Weibull curves, $S(t)=\exp\{-(t/s)^\kappa\}$, and found $\kappa<1$: \citet{hamilton2026peto} reports a shape near $0.37$ for humans and between $0.6$ and $0.9$ for state-of-the-art models, and \citet{ord2026hazard} concludes that, ``contrary to my earlier hypothesis, AI agents probably don't have a constant hazard rate / half-life. Instead their hazard rates systematically decline as the task goes on.''

This reading conflates two things that a cross-section of tasks cannot separate: how the hazard evolves within a task, and how tasks differ from one another. The point is classical in demography and survival analysis \citep{vaupel1979impact,vaupel1985ruses}, and \citet{ord2025halflife} himself notes in a footnote that averaging exponential curves with different rates would make the aggregate hazard look as if it declines. We make it precise for horizon curves and then ask what can be learned instead.

\paragraph{Contributions.}
\begin{enumerate}
\item \emph{The cross-section bounds the within-task shape from below only} (\cref{sec:cross}). A log-logistic horizon curve with $\beta\le1$ is a mixture of constant hazards, and in a proportional-hazards frailty model it can be produced by a within-task Weibull shape $k$ if and only if $k\ge\beta$ (\cref{prop:cm,prop:kbound}). Both statements are short consequences of classical results \citep{schilling2012bernstein,hougaard1986survival,pillai1990mittag}; their content is that no log-logistic or Weibull horizon curve, however well measured, can rule out a hazard that increases within a task. The qualitative point has also been made informally by \citet{mahendrakar2026three} and in the documentation of \citet{ghosh2026horizonkit}.
\item \emph{Repeated runs identify $k$ inside the model, under conditionally independent runs} (\cref{sec:repeated}). The pass$^2$ curve, $\Pr(\text{two runs both succeed})$, equals the single-run curve shifted by $\theta=1/k$ in $\log_2 t$; with a log-logistic cross-section this is a pure intercept shift in logit. Without the Weibull form, the horizontal distance between the two curves is still identified and lies at or above one whenever the average within-task hazard is non-increasing; when the cross-section is log-logistic, the limit of a simple two-step estimator inherits this bound (\cref{prop:direction}).
\item \emph{What end-of-task data can and cannot see, and what data test the prefix assumption} (\cref{sec:diag}). End-of-task data cannot distinguish a prefix violation from a prefix-invariant model whenever they are compatible with one (\cref{thm:diag,prop:compat,cor:stat}). Outcomes at a fixed checkpoint, on prefix-comparable tasks, test the assumption, and pairing the same instance at two lengths gives an exact paired test of the instance-level null (\cref{thm:test}).
\item \emph{Four preregistered experiments on two open model families} (\cref{sec:experiments}). In all four, average paired accuracy on the common first steps is lower at the longer length, and the paired test, a preregistered primary test in three of them and a component of the primary comparison in the fourth, rejects each time. The length of the lists the task must index is the largest single part of the effect in point estimate on both models (about half of it on the second). In five descriptive comparisons on the first model, end-of-task estimators report steeper shapes than within-run estimators on the same data.
\item \emph{A conditional reanalysis of METR's run-level data} (\cref{sec:metr}), with its assumptions listed next to the results. Like every estimator built on end-of-task data, it measures the diagonal of \cref{sec:diag}; it is a within-task shape only under the prefix assumption.
\item \emph{A map of what breaks the shift} (\cref{app:mechanisms}). Five mechanisms, with what each does to the estimate (a direction where one can be given) and the data feature that can or cannot remove it (\cref{tab:mech}).
\end{enumerate}

\paragraph{Related work.} \citet{nguyen2026estimation} relax the linearity of AI difficulty in $\log$ human time with splines and item-response models, and in one model use a beta-binomial likelihood with a single within-pair correlation parameter, which down-weights AI--task pairs with many runs; they do not study within-task dynamics. \citet{moss2026metr} fits a two-parameter item-response model in which each task has a random difficulty residual around its log human time, and shows that this heterogeneity separates typical from marginal success curves and moves the 80\% horizon by about an order of magnitude. \citet{barry2026modelling} examines how alternative curve families and likelihoods change METR's time-horizon estimates, and uses simulation--extrapolation (SIMEX) to gauge the effect of noise in human task lengths. \citet{nguyen2026estimation} also observe that 83\% of METR's run vectors are unanimous against about 60\% predicted under independence, which is the information we use; \citet{trantruong2026measuring} treat the gap between pass$^k$ and the $k$-th power of the single-run rate as a diagnostic of hidden heterogeneity. \citet{merlet2026reliability} fit pass$^4$ curves to the same run-level data and report that the 50\% horizon at which four runs all succeed is, in geometric mean over models, 2.33 (TH1.0) and 2.39 (TH1.1) times shorter than the single-run horizon; in $\cM$ this ratio is $4^{1/k}$ by \cref{prop:shift}, so their averages correspond to $1/k\approx0.61$ and $0.63$ (their samples and pass$^4$ fits differ from ours). They read the ratio as a reliability gap and do not model within-task dynamics. \citet{mahendrakar2026three} argues that the constant-hazard, step-level and log-logistic accounts describe different levels of the same data and recommends reporting per-task success distributions from repeated attempts and length-controlled versions of realistic tasks (\cref{sec:experiments} builds such a design on a synthetic task and uses it to test the prefix assumption), and the documentation of \citet{ghosh2026horizonkit} states that pooling or replication, not a better fit, restores identification. \citet[Proposition~1]{anon2026pooling} show, for a finite design, that repeated runs enter the likelihood only through the Laplace transform of the frailty at multiples of the baseline hazard. We work with the population curves and derive the shift identity, the doubling-map criterion for the direction of the hazard, and the conditions under which the baseline is identified (\cref{rem:three}).

In econometrics the same confound of true and spurious duration dependence in mixed proportional hazards models is resolved with observed covariates in single-spell data \citep{elbers1982true,heckman1984identifiability} and, without moment or tail conditions on the mixing distribution, with multiple spells \citep{honore1993identification}. Repeated runs of a task at one length observe only the diagonal of the bivariate survival function, because the two runs share both the frailty and the length, which is why a moment condition returns in \cref{rem:three}.

Controlled experiments that score errors step by step or vary the length of the same task \citep{sinha2026illusion,serval2026context,hou2025sustained} address within-task dynamics directly; our analysis says what horizon data alone can and cannot add to them.

\paragraph{Concurrent work.} \citet{anon2026pooling}, an anonymous submission under review that we learned of after posting the first version of this paper, reach a closely related conclusion on METR's data independently. They use repeated runs to fit the distribution of task frailty nonparametrically and find corrected Weibull shapes above one. They also show that, with difficulty scaling as a power of length, cross-task data measure $k+\beta$, and they state that their Assumption~1 (length-invariant difficulty) cannot be checked from data of that shape. \Cref{sec:diag} generalises that observation beyond the power law and beyond any parametric baseline, and says what data can check it; \cref{sec:experiments} checks the corresponding prefix invariance on generated tasks, not their assumption on METR's data. Within a parametric family a gradient that is not a power law can still show up as lack of fit, as \citet{anon2026pooling} note; \cref{thm:diag} shows that this detectability comes from the parametric family alone. \citet{li2026fragility} find that composing independent sub-tasks makes each sub-task less accurate, and attribute this to implicit reasoning budgets. \citet{chen2026multiinstance} find that the number of instances matters more than context length, and \citet{chen2026order} report order effects across instances. The phenomenon we find, steps that a model can do becoming harder inside a longer task, is of the same kind. Our independent-expression arm shows it too, so we do not claim that it is specific to dependent chains. What is new here is the test of the assumption on which a survival reading of end-of-task data rests, and an exact version of it.

\section{Setup, and what a single run per task cannot identify}\label{sec:cross}

Fix an agent (model plus scaffold). Task $i$ has human time $t_i>0$ (minutes) and is run $n_i$ times with outcomes $Y_{ia}\in\{0,1\}$; write $y_i=\sum_a Y_{ia}$ and $x=\log_2 t$.

\begin{assumption}[A0: conditionally independent runs]\label{ass:a0}
Given task $i$, the runs are independent with common success probability $p_i$.
\end{assumption}

Define the single-run curve $S(t)=\E[p_i\mid t_i=t]$ and the moment curves $m_r(t)=\E[p_i^r\mid t_i=t]$. Under A0, $\binom{y_i}{r}/\binom{n_i}{r}$ is unbiased for $p_i^r$ when $r\le n_i$, so with many tasks every $m_r$ with $r\le\min_i n_i$ is identified pointwise; $m_r$ is the pass$^r$ curve of agent evaluation \citep{yao2024taubench}, and this is its usual unbiased estimator. METR's fit is $\logit S(t)=\alpha+\gamma x$ with $\gamma<0$, the log-logistic curve $S(t)=1/\{1+(t/h)^\beta\}$ with $\beta=-\gamma/\ln2$ and 50\% horizon $h$.

In survival analysis a \emph{frailty} is an unobserved multiplier on a unit's hazard; here it plays the role of task difficulty, shared by every run of a task.

\begin{definition}[Frailty model $\cM$]\label[definition]{def:model}
Task $i$ carries a cumulative hazard $Z_i H_0(t)$ with $H_0(t)=(t/s)^k$, where $Z_i\ge0$ has an arbitrary distribution and is independent of $t_i$ (so the distribution of difficulty does not change with task length); thus $p_i=\exp\{-Z_iH_0(t_i)\}$ and $S(t)=L_Z(H_0(t))$, with $L_Z$ the Laplace transform of $Z$. Read as a within-task shape, which needs the prefix assumption below, $k>1$ is a hazard that rises with progress through a task and $k<1$ one that falls.
\end{definition}

Reading $k$ as a within-task shape requires a \emph{prefix assumption}: a task that takes a human $t$ minutes behaves like the same hazard process run for $t$ minutes. Horizon data never observe a task at two lengths. \Cref{sec:diag} shows what this costs: whenever such data are compatible with the prefix assumption, they cannot detect its failure, and distinguishing such a violation takes additional observations, such as outcomes at a common checkpoint; \cref{sec:experiments} does so.

\begin{proposition}[Mixtures of constant hazards; classical]\label[proposition]{prop:cm}
If $0<\beta\le1$, the log-logistic curve $S(t)=1/\{1+(t/h)^\beta\}$ is completely monotone on $(0,\infty)$, hence $S(t)=\int_0^\infty e^{-\lambda t}\,\mu(d\lambda)$ for a probability measure $\mu$ on $(0,\infty)$: every task could have a constant hazard, with rates differing across tasks. If $\beta>1$ the hazard of $S$ first rises and then falls, so $S$ is not such a mixture. The same holds for the Weibull curve with $\kappa\le1$.
\end{proposition}

\begin{proposition}[The cross-section bounds $k$ from below]\label[proposition]{prop:kbound}
The log-logistic curve with parameter $\beta$ is generated by some member of $\cM$ with within-task shape $k$ if and only if $k\ge\beta$. The Weibull curve with parameter $\kappa$ is generated if and only if $k\ge\kappa$.
\end{proposition}

Proofs are in \cref{app:proofs}. \Cref{prop:kbound} follows from classical facts: in the log-logistic case the required frailty is a scaled Mittag-Leffler variable of index $\beta/k\le1$ \citep{pillai1990mittag}, in the Weibull case a positive stable variable of index $\kappa/k\le1$ \citep{hougaard1986survival}. Its practical content is a one-line check: since $t_{80}/t_{50}=4^{-1/\beta}$ for the log-logistic curve, $\beta\le1$ if and only if $t_{80}/t_{50}\le1/4$. If $\beta<1$, constant, decreasing and increasing within-task hazards ($k=1$, $\beta\le k<1$, $k>1$) are all compatible with the curve; if $\beta=1$, constant and increasing hazards are; if $\beta>1$, only increasing ones. A Weibull fit with $\kappa<1$ shows only that $k\ge\kappa$. As an illustration, data simulated with a constant within-task hazard ($k=1$) and Mittag-Leffler heterogeneity give a single-run Weibull fit of $\hat\kappa=0.532$ (\cref{app:prov}).

\section{Repeated runs restore identification}\label{sec:repeated}

\begin{proposition}[Moment shift]\label[proposition]{prop:shift}
In $\cM$ under A0, $m_r(t)=S(r^{1/k}t)$ for every $r\ge1$. In particular, with $\theta:=1/k$,
\begin{equation}\label{eq:T}
m_2(x)=S(x+\theta)\qquad\text{on the $\log_2$ scale.}\tag{T}
\end{equation}
\end{proposition}
\begin{proof}
$m_r(t)=\E e^{-rZH_0(t)}=L_Z(rH_0(t))=L_Z(H_0(r^{1/k}t))=S(r^{1/k}t)$.
\end{proof}

\begin{corollary}[Logit intercept shift]\label[corollary]{cor:logit}
If $\logit S(x)=\alpha+\gamma x$, then $\logit m_2(x)=\alpha+\gamma x+\gamma\theta$: the two curves have the same slope and the intercept moves by $\delta=\gamma\theta$. Constant hazard ($k=1$) is $\delta=\gamma$; \cref{prop:kbound} gives $\delta\in[-\ln 2,0)$.
\end{corollary}

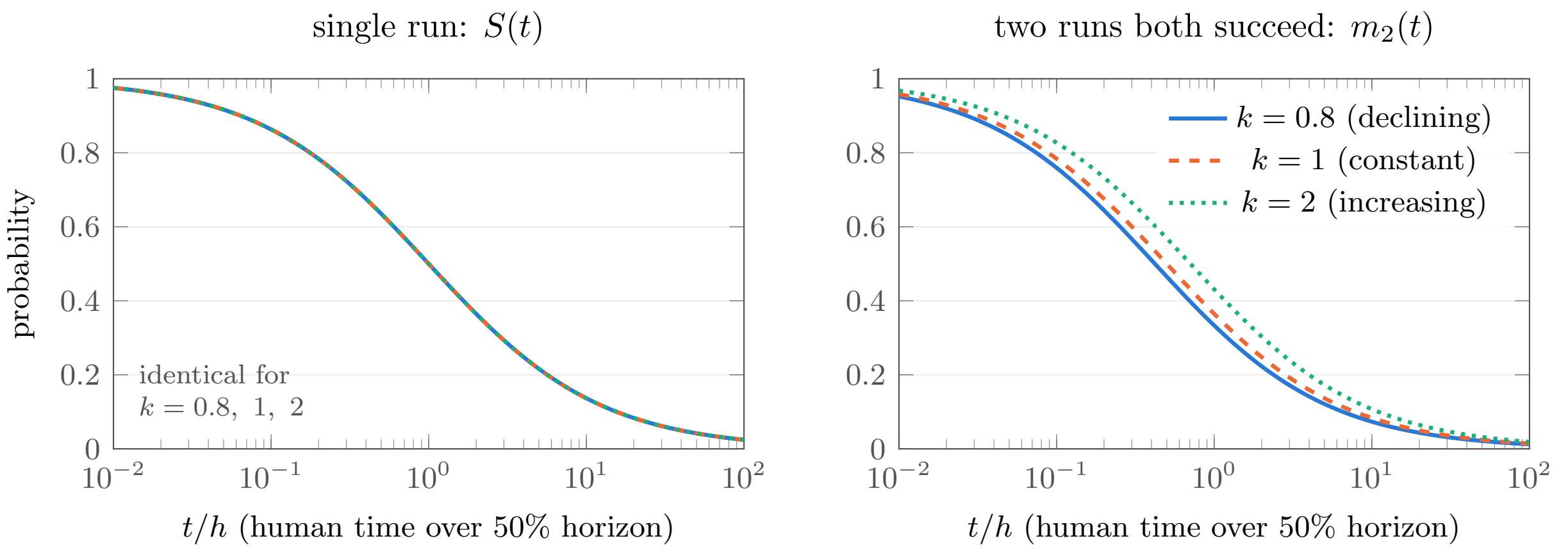
\begin{figure}[t]
\centering
\begin{tikzpicture}
\begin{groupplot}[
  group style={group size=2 by 1, horizontal sep=1.6cm},
  width=0.49\linewidth, height=5.4cm,
  xmode=log, xmin=0.01, xmax=100, ymin=0, ymax=1,
  domain=0.01:100, samples=160,
  xlabel={$t/h$ (human time over 50\% horizon)},
  tick label style={font=\footnotesize, color=cInk}, label style={font=\footnotesize},
  title style={font=\small},
  axis line style={cInk}, major tick style={cInk},
  ymajorgrids, grid style={black!8},
]
\nextgroupplot[title={single run: $S(t)$}, ylabel={probability}]
\addplot[cBlue, line width=1.2pt] {1/(1+x^0.8)};
\addplot[cOrange, line width=1.2pt, dashed] {1/(1+x^0.8)};
\addplot[cAqua, line width=1.2pt, dotted] {1/(1+x^0.8)};
\node[font=\scriptsize, color=cInk, anchor=south west, align=left] at (axis cs:0.012,0.04) {identical for\\ $k=0.8,\ 1,\ 2$};
\nextgroupplot[title={two runs both succeed: $m_2(t)$},
  legend style={font=\footnotesize, at={(0.97,0.97)}, anchor=north east, draw=none, fill=white, fill opacity=0.85, text opacity=1}]
\addplot[cBlue, line width=1.2pt] {1/(1+2^(0.8/0.8)*x^0.8)};
\addlegendentry{$k=0.8$ (declining)}
\addplot[cOrange, line width=1.2pt, dashed] {1/(1+2^(0.8/1)*x^0.8)};
\addlegendentry{$k=1$ (constant)}
\addplot[cAqua, line width=1.2pt, dotted] {1/(1+2^(0.8/2)*x^0.8)};
\addlegendentry{$k=2$ (increasing)}
\end{groupplot}
\end{tikzpicture}
\caption{Three within-task hazard shapes with the same horizon curve. Left: the single-run curve $S(t)=1/\{1+(t/h)^{0.8}\}$, produced in $\cM$ by $k=0.8$, $1$ and $2$ with the frailties of \cref{prop:kbound}; no cross-sectional data can tell them apart. Right: the probability that two runs of the same task both succeed, $m_2(t)=S(2^{1/k}t)=1/\{1+2^{0.8/k}(t/h)^{0.8}\}$ (\cref{prop:shift}), separates them. Analytic curves; no data.}\label{fig:theory}
\end{figure}

So in $\cM$ the agreement between runs of the same task carries the information that the cross-section lacks (\cref{fig:theory}). If every task were always or never solved, $m_2=S$, the limit $\theta\to0$ ($k\to\infty$). If all tasks of a given length had the same success probability (constant frailty), $m_2=S^2$; within $\cM$ this forces a Weibull cross-section with $\kappa=k$, so it is not attainable with a log-logistic cross-section, for which $\theta$ ranges over $(0,1/\beta]$ and $\theta=1/\beta$ already requires an exponential, non-degenerate frailty. \Cref{prop:shift} is the Laplace-transform calculation that underlies shared-frailty models \citep{oakes1989bivariate}. Within $\cM$ the correlation of two runs at a given length, $(m_2-S^2)/\{S(1-S)\}$, equals $(2-q+u)/(1+qu)$ with $u=(t/h)^\beta$ and $q=2^{\beta/k}$ for a log-logistic cross-section, which increases with $u$; so a beta-binomial model with one correlation at every length is not a special case of $\cM$ except when $m_2=S$. This correlation is marginal over the frailty at a given length; the overdispersion parameter of \citet{nguyen2026estimation} is conditional on a task's latent difficulty, so this is not a test of their model. From the population curves $S$, $m_2$ and $m_3$ the baseline $H_0$ itself is identified up to scale without assuming the Weibull form, and from $S$ and $m_2$ alone it is identified when $\E Z<\infty$ but not, in general, otherwise (\cref{rem:three}); these are frailty versions of known results on diagonal sections of Archimedean copulas. Within $\cM$, a log-logistic cross-section has $\E Z<\infty$ only at $k=\beta$, so in the case of most interest here $m_2$ alone does not pin down $H_0$ (\cref{rem:three}(c)). It does pin down the direction that matters, as the next result shows after the estimator is introduced.

\paragraph{Estimator.} We use the estimator (E):
(i) fractional-logit regression \citep{papke1996econometric} of $\hat p_i=y_i/n_i$ on $(1,x_i)$ with weights $w_i$, giving $(\hat\alpha,\hat\gamma)$;
(ii) intercept-only fractional-logit regression of $\hmtwo_{2,i}=y_i(y_i-1)/\{n_i(n_i-1)\}$ with offset $\hat\alpha+\hat\gamma x_i$, giving $\hat\delta$;
(iii) $\htheta=\hat\delta/\hat\gamma$.
Its \emph{population limit}, the pseudo-true value in the sense of \citet{white1982maximum}, replaces $\hat p_i$ and $\hmtwo_{2,i}$ by their expectations, and the first-step estimates in the offset by their limits, and solves the same score equations on the fixed design; because both scores are linear in the responses, replacing the responses by their expectations is exact. Under $\cM$, A0, (T) and a log-logistic cross-section, the limit is $\theta$. In general it is not, and the next result says how far it can move and why.

\begin{proposition}[Representation of the estimator]\label[proposition]{prop:repr}
Fix the design $(x_i,w_i)$ with $w_i>0$, suppose $\hat\gamma<0$, write $\hS(x)=\expit(\hat\alpha+\hat\gamma x)$ and let $m_{2,i}\in[0,1]$ be arbitrary (no shift identity is assumed). Define $G(\vartheta)=\sum_iw_i\{m_{2,i}-\hS(x_i+\vartheta)\}$.
\begin{enumerate}
\item Step (ii) has a finite solution if and only if $0<\sum_iw_im_{2,i}<\sum_iw_i$; then $\htheta$ is the unique root of $G$, which is strictly increasing.
\item For every reference value $\theta$, $\operatorname{sign}(\htheta-\theta)=-\operatorname{sign}G(\theta)$, and $|\htheta-\theta|=|G(\theta)|/G'(\vartheta^*)$ for some $\vartheta^*$ between $\theta$ and $\htheta$.
\item If step (i) targets $S$ and is taken at its population limit (responses replaced by their means $S(x_i)$), and $e:=S-\hS$,
\begin{equation}\label{eq:decomp}
G(\theta)=\underbrace{\sum_iw_i\{m_{2,i}-S(x_i+\theta)\}}_{\text{violation of (T)}}+\underbrace{\sum_iw_i\{e(x_i+\theta)-e(x_i)\}}_{\text{projection residual}}.
\end{equation}
\end{enumerate}
\end{proposition}

\begin{proof}
With $\delta=\hat\gamma\vartheta$, the score of step (ii) is $\sum_iw_i\{m_{2,i}-\expit(\hat\alpha+\hat\gamma x_i+\delta)\}=G(\vartheta)$. Its derivative is $-\hat\gamma\sum_iw_i\hS(1-\hS)(x_i+\vartheta)>0$, and $G\to\sum w_im_{2,i}-\sum w_i$ as $\vartheta\to-\infty$ and $G\to\sum w_im_{2,i}$ as $\vartheta\to\infty$; this gives 1, and 2 is the mean value theorem. For 3, add and subtract $\sum w_iS(x_i+\theta)$ and use the intercept score of step (i), $\sum_iw_ie(x_i)=0$.
\end{proof}

The projection residual vanishes when step (i) is exact on $\{x_i\}\cup\{x_i+\theta\}$, in particular when $S$ is log-logistic. Otherwise $e(x_i+\theta)$ is evaluated off the design, nothing in step (i) controls it, and we have no general bound.

\begin{proposition}[Direction without the Weibull form]\label[proposition]{prop:direction}
Let $p_i=\exp\{-Z_iH_0(t_i)\}$ under A0, with $Z\indep t$, $\Pr(Z>0)>0$, and $H_0$ continuous and strictly increasing with $H_0(0+)=0$ and $H_0(\infty)=\infty$; nothing else is assumed about $H_0$ or the law of $Z$. Define the doubling map $\psi_2:=S^{-1}\circ m_2$ and its horizontal distance $\theta(x):=\log_2\psi_2(2^x)-x$.
\begin{enumerate}
\item $H_0(\psi_2(t))=2H_0(t)$; hence $\theta(\cdot)$ is identified from $S$ and $m_2$ with no moment condition, and it is the constant $1/k$ when $H_0(t)=(t/s)^k$.
\item If the average hazard $H_0(t)/t$ is non-increasing (DFRA, decreasing failure rate on average: the hazard accumulated per unit of time does not grow), then $\theta(x)\ge1$ for every $x$; if it is non-decreasing (IFRA), then $\theta(x)\le1$.
\item If $H_0$ is concave, $S$ has a non-increasing hazard.
\item In \cref{prop:repr} take $m_{2,i}=m_2(x_i)$ and let step (i) target $S$. Under DFRA, $G(1)$ is at most the projection residual at $\theta=1$. If step (i) is exact on $\{x_i\}\cup\{x_i+1\}$, the population limit of $\htheta$ is therefore at least one.
\end{enumerate}
\end{proposition}
\begin{proof}
1: $L_Z$ is strictly decreasing because $\Pr(Z>0)>0$, so $S$ maps $(0,\infty)$ onto $(\Pr(Z=0),1)$, which contains every value of $m_2$; $S(\psi_2(t))=m_2(t)$ reads $L_Z(H_0(\psi_2(t)))=L_Z(2H_0(t))$. 2: under DFRA, $H_0(2t)\le2H_0(t)=H_0(\psi_2(t))$, so $2t\le\psi_2(t)$; IFRA is symmetric. 3: for concave $H_0$ each $t\mapsto e^{-zH_0(t)}$ is log-convex, and a mixture of log-convex functions is log-convex, so $-\log S$ is concave and its derivative, which exists almost everywhere, is non-increasing. 4: the existence condition of \cref{prop:repr}(1) holds because $0<m_2<1$; $m_2(x_i)=S(x_i+\theta(x_i))\le S(x_i+1)$ by part 2, so the first sum in \eqref{eq:decomp} is non-positive at $\theta=1$; with an exact first step the projection residual is zero, and \cref{prop:repr}(2) gives $\htheta\ge1$.
\end{proof}

So a confidence interval for the population limit of $\htheta$ lying below one is evidence against a non-increasing average within-task hazard (in particular against a constant one) without the Weibull form. It still rests on
\begin{itemize}\setlength{\itemsep}{0pt}\setlength{\parskip}{0pt}
\item A0 or the cross-batch conditions;
\item $Z\indep t$ and the selection and composition conditions of \cref{sec:mechanisms};
\item the log-logistic cross-section, which is an assumption about the observable curve;
\item the prefix assumption needed to read $H_0$ as a within-task hazard.
\end{itemize}
It is not evidence that the average hazard increases everywhere, since $\theta(x)$ may cross one; conversely, a limit of $\htheta$ at or above one does not establish DFRA, because $\theta(x)$ can fall below one on part of the range while the weighted estimator stays above one. Every inference here is one-sided. Part 3 gives a cross-sectional counterpart: a log-logistic curve with $\beta>1$, whose hazard increases near zero, already excludes a concave baseline. A concave $H_0$ with $H_0(0+)=0$ is DFRA, so this is consistent with part 2, and since the concave class contains the linear baseline it is stronger than the exclusion of $k=1$ in \cref{prop:cm}; it does not exclude $k>1$. The criterion uses only the diagonal of the bivariate survival function of two runs, the part that repeated runs observe; ageing of exchangeable lifetimes through that function is studied in \citet{nappo2020relations} and the references there.

\section{End-of-task data, the diagonal, and testing the prefix assumption}\label{sec:diag}

The estimators of \cref{sec:repeated} use one outcome per run: whether the task was completed. This section asks what such data can say about the prefix assumption of \cref{sec:cross}, and what data can test it. The answer has three parts. End-of-task data can rule out the assumption only in special cases; whenever they are compatible with it, they cannot tell it from its failure (\cref{thm:diag,prop:compat}). Recording, in addition, whether a run is still correct at a fixed checkpoint makes the assumption testable, and running the same instance at two lengths makes it testable instance by instance, with an exact test (\cref{thm:test}).

\paragraph{Setting.} Progress is counted in steps $s=1,2,\dots$, and the design uses finitely many task lengths $\mathcal C=\{c_1<\dots<c_L\}$. A base instance $X_i$ is drawn from a population, and for each $c\in\mathcal C$ a version $A_c(X_i)$ of length $c$ is defined in advance. We call the versions \emph{prefix-comparable} if, on every common step $s\le\min(c,c')$, $A_c(X_i)$ and $A_{c'}(X_i)$ have the same content (or a correspondence fixed in advance), the same gold answer and the same scoring, and if which versions an instance receives does not depend on its potential outcomes. This is a property of the design. Equal success rates on the common steps are what is to be tested, so they are not part of the definition. Let $T$ be the first erroneous step of a run, and
\[
P_i(s\mid c)=\Pr(T>s\mid i,c),\qquad s=0,1,\dots,c,
\]
so $P_i(0\mid c)=1$ and $s\mapsto P_i(s\mid c)$ is non-increasing. The \emph{diagonal} is $D_i(c)=P_i(c\mid c)$, the probability that a run of the length-$c$ version gets all $c$ steps right. End-of-task data (horizon data, and the cross-sectional fits of \cref{sec:repeated} and of \citealp{anon2026pooling}) observe only the diagonal, at the length each instance happens to have. In horizon data each task has one length, and different lengths are different tasks. The results below apply there to the laws of the diagonal across lengths, under the assumption that the task populations at different lengths are exchangeable, which in $\cM$ is $Z\indep t$.

\begin{assumption}[A0$^{+}$: independent runs across lengths]\label{ass:a0plus}
Given instance $i$, all runs of all its versions are mutually independent, each run of version $c$ has $\Pr(T>s)=P_i(s\mid c)$ for every $s$, and runs of different instances are independent.
\end{assumption}

A0 (\cref{ass:a0}) requires independence only within a version. That suffices for the moment estimators below, but not for the paired test of \cref{thm:test}(3). Two arms can each be binomial and satisfy A0 while their counts are dependent in a way that makes the difference asymmetric (\cref{app:proofs}).

Two versions of the prefix assumption are relevant:
\begin{itemize}\setlength{\itemsep}{0pt}
\item $H_{\mathrm{inst}}$: $P_i(s\mid c)=P_i(s\mid c')$ for all $s\le\min(c,c')$, almost surely;
\item $H_{\mathrm{law}}$: for $c<c'$, the vector $(P_i(s\mid c'))_{s\le c}$ has the same distribution as $(P_i(s\mid c))_{s\le c}$.
\end{itemize}
$H_{\mathrm{inst}}$ implies $H_{\mathrm{law}}$. In a frailty model with a common baseline, $P_i(s\mid c)=\exp\{-Z_iH_0(s)\}$ where only the law of $Z_i$ may depend on $c$, $H_{\mathrm{law}}$ is equivalent to that law not depending on $c$ as soon as $0<H_0(s)<\infty$ at some common step. This is the length-invariant difficulty of $\cM$ and Assumption~1 of \citet{anon2026pooling}. If the baseline may also depend on $c$, rejecting $H_{\mathrm{law}}$ rejects the two restrictions jointly.

\Cref{tab:datatypes} summarises what follows.

\begin{table}[ht]
\centering\small
\caption{What each kind of data can identify or test about the prefix assumption ($H_{\mathrm{law}}$: same distribution of early-step success across lengths; $H_{\mathrm{inst}}$: same early-step success for every instance).}\label{tab:datatypes}
\begin{tabularx}{\linewidth}{@{}>{\raggedright\arraybackslash}p{0.27\linewidth}>{\raggedright\arraybackslash}X>{\raggedright\arraybackslash}X@{}}
\toprule
Data & What it can do & Result \\
\midrule
End-of-task outcomes (horizon data) & Refute the assumption only through order restrictions on moments; otherwise indistinguishable from a prefix-invariant model, whatever the estimator & \Cref{thm:diag}, \cref{prop:compat,cor:stat} \\
Checkpoint outcomes, one length per instance & Test $H_{\mathrm{law}}$; cannot separate $H_{\mathrm{inst}}$ from $H_{\mathrm{law}}$ unless the sign of the effect is restricted & \Cref{thm:test}(1)--(2) \\
Checkpoint outcomes, same instance at two lengths & Exact test of $H_{\mathrm{inst}}$ at a checkpoint; identify $V_j$, which is zero iff $H_{\mathrm{inst}}$ holds there & \Cref{thm:test}(3)--(4) \\
Checkpoints within one run & Test the direction of the within-run hazard at a given length, with no prefix assumption & \Cref{prop:wd} \\
\bottomrule
\end{tabularx}
\end{table}

\subsection{What end-of-task data cannot see}

\begin{lemma}[Common-quantile coupling; \citealp{dhaene2002comonotonicity}]\label[lemma]{lem:coupling}
Let $\{\nu_c\}_{c\in\mathcal C}$ be probability measures on $[0,1]$. There are random variables $(Q(c))_{c\in\mathcal C}$ with $Q(c)\sim\nu_c$ and $Q$ almost surely non-increasing in $c$ if and only if $\nu_{c'}\le_{\mathrm{st}}\nu_c$ whenever $c<c'$. One construction is $Q(c)=\inf\{x:F_c(x)\ge U\}$ with a single $U\sim\mathrm{Unif}(0,1)$.
\end{lemma}

The idea is simple. End-of-task data see, for each length $c$, only whether all $c$ steps were done right. If these probabilities fall with $c$, one can always draw a single survival curve through them, so a model in which every task is a prefix of the next fits the data exactly, whatever really happens on the early steps of long tasks.

\begin{theorem}[Twins of a single instance]\label{thm:diag}
For an instance $i$ the following are equivalent:
\begin{enumerate}\setlength{\itemsep}{0pt}
\item $c\mapsto D_i(c)$ is non-increasing on $\mathcal C$;
\item there are survival paths $P'_i(\cdot\mid c)$, $c\in\mathcal C$, that satisfy $H_{\mathrm{inst}}$ and have $P'_i(c\mid c)=D_i(c)$ for every $c\in\mathcal C$.
\end{enumerate}
In particular both hold if lengthening never makes a common step easier, that is if $P_i(s\mid c')\le P_i(s\mid c)$ for all $s\le c<c'$.
\end{theorem}

\begin{proof}
If (2) holds, the paths share one survival function $Q_i(s)=P'_i(s\mid c)$, which is non-increasing in $s$, and $D_i(c)=Q_i(c)$. Conversely, given (1), set $Q_i(0)=1$, $Q_i(c)=D_i(c)$ on $\mathcal C$ and $Q_i(s)=D_i(c_k)$ for $c_{k-1}<s<c_k$ (with $c_0=0$), and let $P'_i(s\mid c)=Q_i(s)$. For the last claim, $D_i(c')=P_i(c'\mid c')\le P_i(c\mid c')\le P_i(c\mid c)=D_i(c)$: the first inequality holds because survival paths are non-increasing, the second by assumption.
\end{proof}

The sufficient condition is not necessary. If $P_i(1\mid1)=0.7$, $P_i(1\mid2)=0.9$ and $P_i(2\mid2)=0.6$, lengthening makes the first step easier, yet the diagonal falls and a twin exists. The twin is a construction, not a claim about the truth: its interpolation between design lengths is arbitrary.

If every instance has a non-increasing diagonal, the twin model and the true model generate end-of-task data with the same joint law. This holds under any design in which lengths and run counts depend only on instances, baseline covariates and external randomisation, with the same conditional law of runs given the design and the instance, and with independent clusters of runs. The observation equivalence then holds for all run counts at once. With a fixed number of runs per instance, compatibility is weaker still.

\begin{proposition}[Compatibility at a fixed number of runs]\label[proposition]{prop:compat}
Let each instance be observed at one length only, assigned at random, let every instance be run $n$ times, and let $\mu_c$ be the law of $D_i(c)$. The end-of-task data are determined by the moments $m_{c,r}=\int p^r\,d\mu_c(p)$, $r=1,\dots,n$ \citep{wood1999binomial}. A model satisfying $H_{\mathrm{inst}}$ and generating the same end-of-task data exists if and only if there are probability measures $\nu_c$ on $[0,1]$ with $\int p^r\,d\nu_c=m_{c,r}$ for $r\le n$ and $\nu_{c'}\le_{\mathrm{st}}\nu_c$ whenever $c<c'$.
\end{proposition}

\begin{proof}
Given $C=c$, the number of successes $K$ has $\Pr(K=k)=\binom nk\int p^k(1-p)^{n-k}\,d\mu_c$, a linear function of $m_{c,1},\dots,m_{c,n}$, and conversely $m_{c,r}=\E[(K)_r\mid C=c]/(n)_r$ with $(x)_r$ the falling factorial. Necessity: take $\nu_c$ the law of the twin's diagonal, which is non-increasing in $c$. Sufficiency: apply \cref{lem:coupling} to $\{\nu_c\}$ and extend between design lengths as in \cref{thm:diag}.
\end{proof}

So end-of-task data can refute the prefix assumption, but only through order restrictions on moments. A longer task with a higher success rate refutes it; so do some violations visible only in higher moments. For example, $\mu_{24}=\tfrac12(\delta_0+\delta_1)$ and $\mu_{48}=\delta_{1/2}$ have the same mean. With $n=1$ they produce the same data, but with $n\ge2$ their second moments, $\tfrac12$ and $\tfrac14$, cannot be reconciled with ordered laws of equal mean. What end-of-task data cannot do is detect a violation that leaves them compatible. A violation in which longer tasks make their early steps harder typically does leave them compatible, since it only steepens the diagonal.

\begin{corollary}[End-of-task statistics]\label[corollary]{cor:stat}
Under the equivalences above, every statistic computed from end-of-task data has the same law under the true model and under a prefix-invariant twin, and the same limit if it has one. This covers $\htheta$ of \cref{sec:repeated} and the cross-sectional fits of \citet{anon2026pooling} (naive, gamma, log-normal and NPMLE). It does not by itself say that the limit is the shape of any path. Reading it as a within-task shape needs, in addition, a model in which the estimator is consistent for that shape; misspecified fits are projections (\cref{prop:repr}).
\end{corollary}

The construction also covers shared-frailty models and the power-law case of \citet[App.~B.3]{anon2026pooling}; \cref{rem:frailty} in \cref{app:proofs} gives an exact split of the diagonal's slope into a within-run change and a length effect, and relates the result to duration analysis.

\subsection{What checkpoint data can test}

Fix a checkpoint $j$ and two lengths $c<c'$ with $j\le c$, chosen in advance. Write $K_i^{c}(j)$ for the number of the $n$ runs of version $A_c(X_i)$ whose first $j$ steps are correct.

\begin{theorem}[Testing the prefix assumption]\label{thm:test}
Let the versions be prefix-comparable.
\begin{enumerate}\setlength{\itemsep}{2pt}
\item \emph{Unpaired data test $H_{\mathrm{law}}$.} If each instance is run at one length with first-error steps recorded, the data identify, for each $c$, the mixed moments of total degree at most $n$ of $(P_i(s\mid c))_{s\le c}$. Under $H_{\mathrm{law}}$ the laws of first-error counts, truncated at step $c$, coincide across $c$ and $c'$. To test at a single checkpoint $j$ it suffices to record whether each run is still correct at step $j$.
\item \emph{Unpaired data cannot separate $H_{\mathrm{inst}}$ from $H_{\mathrm{law}}$} unless the sign of the length effect is restricted. There are models satisfying $H_{\mathrm{law}}$ but not $H_{\mathrm{inst}}$ that are observationally equivalent, for every $n$, to models satisfying $H_{\mathrm{inst}}$. If the length effect is known to be weakly harmful for every instance, the unpaired first moments already characterise $H_{\mathrm{inst}}$ at $j$.
\item \emph{Paired data give an exact test of $H_{\mathrm{inst}}$ at $j$.} Under A0$^{+}$ and the null $P_i(j\mid c)=P_i(j\mid c')$ a.s., the differences $\Delta_i=\{K_i^{c}(j)-K_i^{c'}(j)\}/n$ are independent across instances and symmetric about zero. Conditionally on $(|\Delta_i|)_i$, the sign-flip test of $\bar\Delta$ has level at most $\alpha$ in finite samples.
\item \emph{Paired data identify a measure of instance-level change.} Under A0$^{+}$ and $n\ge2$, with $p_i=P_i(j\mid c)$ and $q_i=P_i(j\mid c')$,
\[
V_j:=\E(p_i-q_i)^2=\E\!\left[\frac{(K_i^{c})_2+(K_i^{c'})_2}{(n)_2}-\frac{2K_i^{c}K_i^{c'}}{n^2}\right],
\]
and $V_j=0$ if and only if $H_{\mathrm{inst}}$ holds at $j$.
\end{enumerate}
\end{theorem}

\begin{proof}
(1) Let $q_s=P_i(s-1\mid c)-P_i(s\mid c)$ for $s\le c$ and $q_{c+1}=P_i(c\mid c)$. Given the instance, the counts $N$ of first errors at each step (and of runs with $T>c$) are multinomial$(n,q)$, and $\E[\prod_s(N_s)_{r_s}\mid C=c]/(n)_{\sum r_s}=\E[\prod_s q_s^{r_s}\mid C=c]$ for $\sum_s r_s\le n$. Since $q$ is an affine function of the path, these are its mixed moments of total degree at most $n$. Under $H_{\mathrm{law}}$ the truncated paths have the same law at $c$ and $c'$, hence so do the truncated counts. (2) Take two paths $f\ne h$ on the common steps and two latent types in equal proportions, with paths $(f,h)$ at length $c$ and $(h,f)$ on the common steps at length $c'$, extended monotonically. Every observed arm sees the same mixture of paths as in the model with paths $(h,f)$ at both lengths, which satisfies $H_{\mathrm{inst}}$. If $P_i(j\mid c)-P_i(j\mid c')\ge0$ a.s., a zero mean forces a zero difference. (3) Under the null and A0$^{+}$, $K_i^{c}(j)$ and $K_i^{c'}(j)$ are independent and identically binomial given $i$, so $\Delta_i$ and $-\Delta_i$ have the same law; conditionally on $|\Delta_i|$ the non-zero signs are independent fair coins. (4) $\E[(K)_2/(n)_2\mid i]$ equals $p_i^2$ (respectively $q_i^2$), and by A0$^{+}$, $\E[K_i^{c}K_i^{c'}\mid i]=n^2p_iq_i$.
\end{proof}

In part (3) the two versions may be worded differently. A rejection then says that these two prompts give the same instance different chances on its first $j$ steps, and it is attributable to length only if length is all they differ in. The checkpoint, the direction and the pair of prompts must be fixed in advance. With Monte Carlo sign flips, reported $p$-values are limited by the number of flips. $V_j$ is not cancelled when length effects of opposite signs average out, as $\bar\Delta$ can be. Swapping the arms within an instance leaves $\hat V_j$, the average over instances of the bracket in part (4), unchanged, so its sign-flip distribution is degenerate; under A0$^{+}$ and the null, conditioning instead on each instance's total $K_i^{c}(j)+K_i^{c'}(j)$ makes the splits between the arms independent and hypergeometric, which gives a finite-sample test of level at most $\alpha$ for any statistic of the paired counts fixed in advance. With a common $n\ge2$, $\hat V_j$ is a positive multiple of the score statistic for the variance of independent mean-zero normal log odds ratios in this conditional model (cf.\ \citealp{lin1997variance}), so the upper-tail test based on it is that score test with exact calibration; it sums squared instance scores where $\bar\Delta$ sums signed ones, as in the burden--SKAT comparison of \citet{lee2012optimal}.

\begin{proposition}[Within-run shapes and length]\label[proposition]{prop:scale}
Suppose $-\log P_i(s\mid c)=Z_iH(s\mid c)$ with $Z_i$ shared across lengths and $0<H(j\mid c)<\infty$. Then
\[
P_i(j\mid c')=P_i(j\mid c)^{H(j\mid c')/H(j\mid c)}.
\]
Call two within-run shapes equal if the cumulative hazards $H(\cdot\mid c)$ and $H(\cdot\mid c')$ differ by a positive factor on the common steps. Shapes are then equal at all lengths if and only if the length effect is of scale type, $H(\cdot\mid c')=a(c,c')H(\cdot\mid c)$. Equality of a fitted $k$ is weaker than this. Without scale type there is no within-run shape that does not depend on length. Ratios of pooled cumulative hazards do not reveal scale type (\cref{app:proofs} gives a gamma-frailty example).
\end{proposition}

\begin{proposition}[A within-run check that needs only A0]\label[proposition]{prop:wd}
Fix $c$ and $2t\le c$, and let $H_D(t)$ be the hypothesis $P_i(2t\mid c)\ge P_i(t\mid c)^2$ a.s. (when cumulative hazards are finite, $\Lambda_i(2t)\le2\Lambda_i(t)$). A non-increasing average hazard $\Lambda_i(s)/s$ implies $H_D(t)$ for every $t$, but not conversely: $\Lambda=(1,1.2,2.4,2.4)$ on $s=1,\dots,4$ satisfies every doubling inequality and is not DFRA. Under $H_D(t)$,
\[
D(t):=\E\{P_i(2t\mid c)-P_i(t\mid c)^2\}\ge0 .
\]
Under A0 with $n\ge2$, the averages over instances of $K_i(2t)/n$ and of $(K_i(t))_2/(n)_2$ are unbiased for the two terms. If $D(t)<0$, a positive fraction of instances accumulate hazard faster on $(t,2t]$ than on $(0,t]$. This says nothing about the instantaneous hazard everywhere, nor about most instances. $H_D(t)$ is rejected when a valid one-sided upper confidence bound for $D(t)$ is below zero.
\end{proposition}

\Cref{prop:wd} is the within-run counterpart of \cref{prop:direction}(2): it compares two checkpoints of the same runs instead of two task lengths, so it needs neither the prefix assumption nor a frailty model. By \cref{prop:scale}, what it measures is the hazard at length $c$, which need not carry over to other lengths.

\section{Testing the prefix assumption on a controlled task}\label{sec:experiments}

\Cref{thm:test} says what data can test the prefix assumption: an outcome recorded at a fixed checkpoint $j$, on tasks whose prefixes are comparable by design across lengths. We built such data and ran four preregistered experiments, three on one model and one on a model from a second family. The preregistrations, frozen code hashes and all deviations are in \cref{app:prereg}, and the results of every preregistered primary check are listed in \cref{tab:primary}.

\paragraph{Task and design.}
\begin{itemize}\setlength{\itemsep}{0pt}
\item \emph{Task.} Nested linear maps, with the prompt of \citet[App.~D]{raju2026model}: given $x_0$ and lists $(a_k)$, $(b_k)$, output $x_k=a_kx_{k-1}+b_k$ for $k=1,\dots,c$. Every step can be checked. We use our own generator, which builds the chain step by step and keeps every value in $\{0,\dots,9\}$. A task of length $24$ is the first $24$ steps of a task of length $48$, with the same initial value and the same coefficients, which makes prefixes comparable by construction.
\item \emph{Model.} Qwen3-8B \citep{yang2025qwen3} with reasoning enabled, run through vLLM \citep{kwon2023efficient} with a 32k-token budget. Runs that hit the budget or give no parsable answer count as failures; on this model the budget was hit by at most two of 800 runs in any arm. Experiment~G uses a second model (below).
\item \emph{Runs.} Each instance is run $8$ times at each length, each request with its own seed. All requests of an experiment are shuffled and submitted as one batch, so lengths are not confounded with batches. The engine shares a prefix cache across requests, so independence of runs across arms (A0$^{+}$) is an assumption that the design makes plausible, not a guarantee.
\item \emph{Experiments.} Each uses 100 instances, drawn afresh, so the four sets of instances are independent.
  \begin{itemize}\setlength{\itemsep}{0pt}
  \item Pilot~A: lengths $\{24,48\}$, temperature $1$.
  \item Replication S1: temperature $1$ at lengths $\{24,48\}$, and the model's recommended reasoning decoding (temperature $0.6$, top-$p$ $0.95$, top-$k$ $20$) at lengths $\{12,24,36,48\}$. One further arm gives lists of length $48$ but asks for the first $24$ values only, with an added sentence ``output only the first 24 values''.
  \item Decomposition S2, recommended decoding, seven arms: the chain at lengths $24$ and $48$; lists of length $48$ or $24$ with the added sentence, each asking for $24$ values; \emph{independent expressions}, which give the same $24$ or $48$ expressions $x_{k-1}a_k+b_k$ with $x_{k-1}$ written out, so that no step uses the previous result and no list needs to be indexed; and steps $25$--$48$ asked as a new task of length $24$ starting from the true $x_{24}$.
  \item Second family G: gpt-oss-20b, the chain at two lengths and the two arms with the added sentence, at lengths $48$ and $96$ chosen by a preregistered calibration rule.
  \end{itemize}
\end{itemize}
Throughout, $\bar\Delta$ is the share of runs correct on the first $j$ steps in the first arm named, minus that in the second, averaged over instances. It is positive when the first arm does better. Block rates are failure rates within a block of six steps, conditional on the earlier steps being correct. A run whose answer cannot be parsed, or whose initial value is wrong, fails every checkpoint; the first block's rate is computed among runs with a correct initial value.

\begin{table}[t]
\centering\small
\caption{Every preregistered primary check. One-sided $p$-values: for $\bar\Delta$, Monte Carlo sign flips ($20{,}000$); for $\hat D$ and S2's main contrast, the instance-bootstrap share on the wrong side of zero ($2{,}000$ and $10{,}000$ resamples), so ``$<5\times10^{-4}$'' is the resolution of $2{,}000$ resamples. ``Holm'' is the adjusted value within the preregistered family; $T{=}1$ is temperature 1 and ``rec.\ decoding'' the recommended decoding. Intervals are instance-bootstrap 95\%.}\label{tab:primary}
\setlength{\tabcolsep}{4pt}\begin{tabularx}{\linewidth}{@{}l>{\raggedright\arraybackslash}Xll@{}}
\toprule
Experiment & Check & Estimate [95\%] & $p$ or share \\
\midrule
Pilot A & (W) $\hat D(24)$ at $c=48$ & $-0.495$ $[-0.552,-0.438]$ & $<5\times10^{-4}$ (Holm) \\ %
 & (P1) $\bar\Delta$, $24$ vs $48$, $j=24$ & $+0.155$ $[+0.107,+0.201]$ & $<10^{-4}$ (Holm) \\ %
S1, $T{=}1$ (gate) & (W) & $-0.525$ $[-0.585,-0.464]$ & $<5\times10^{-4}$ (Holm) \\ %
 & (P1) & $+0.140$ $[+0.098,+0.184]$ & $<10^{-4}$ (Holm) \\ %
S1, rec.\ decoding & (W) & $-0.632$ $[-0.693,-0.569]$ & $<5\times10^{-4}$ (Holm) \\ %
 & (P1) & $+0.128$ $[+0.083,+0.176]$ & $0.0001$ (Holm) \\ %
 & requested length: lists $48$, ask $24$ vs $48$ & $+0.037$ $[+0.003,+0.075]$ & $0.0302$ (Holm) \\ %
 & dose, $24\to36$ & $+0.094$ $[+0.054,+0.136]$ & $<10^{-4}$ \\ %
 & dose, $36\to48$ & $+0.034$ $[-0.003,+0.072]$ & $0.0471$ \\ %
S2 & $\bar\Delta_{\mathrm{dep}}-\bar\Delta_{\mathrm{ind}}$ (dependent minus independent) & $+0.059$ $[+0.004,+0.113]$ & $0.0198$ \\ %
 & \quad same on the logit scale & $-2.354$ $[-4.035,-0.951]$ & --- \\ %
G (gpt-oss-20b) & (W) $\hat D(48)$ at $c=96$ & $-0.132$ $[-0.175,-0.091]$ & $<5\times10^{-4}$ (Holm) \\ %
 & (P1) $\bar\Delta$, $48$ vs $96$, $j=48$ & $+0.315$ $[+0.260,+0.367]$ & $<10^{-4}$ (Holm) \\ %
\bottomrule
\end{tabularx}
\end{table}

\paragraph{The prefix assumption fails, on four independent sets of instances and two models.} The paired check of \cref{thm:test}(3) compares, instance by instance, the share of runs whose first $24$ steps are correct when the task has length $24$ and when it has length $48$ (in G, the first $48$ steps at lengths $48$ and $96$).
\begin{itemize}\setlength{\itemsep}{0pt}
\item Pilot A: $\bar\Delta=+0.155$ ($[+0.107,+0.201]$). %
\item S1, temperature 1: $\bar\Delta=+0.140$ ($[+0.098,+0.184]$). %
\item S1, recommended decoding: $\bar\Delta=+0.128$ ($[+0.083,+0.176]$). %
\item S2: $\bar\Delta=+0.095$ ($[+0.051,+0.141]$), a component of S2's primary check rather than itself a preregistered test. %
\item G (gpt-oss-20b), lengths $48$ and $96$, $j=48$: $\bar\Delta=+0.315$ ($[+0.260,+0.367]$). %
\end{itemize}
In Pilot~A, for example, 697 of 800 runs got the first 24 steps right at length 24 and 573 of 800 at length 48: per instance, 15.5 percentage points more of its runs, or 1.24 more of its eight on average. The difference is positive for 65 instances, zero for 18 and negative for 17, so on average the same instance does worse on the same steps when they are part of a longer task. %

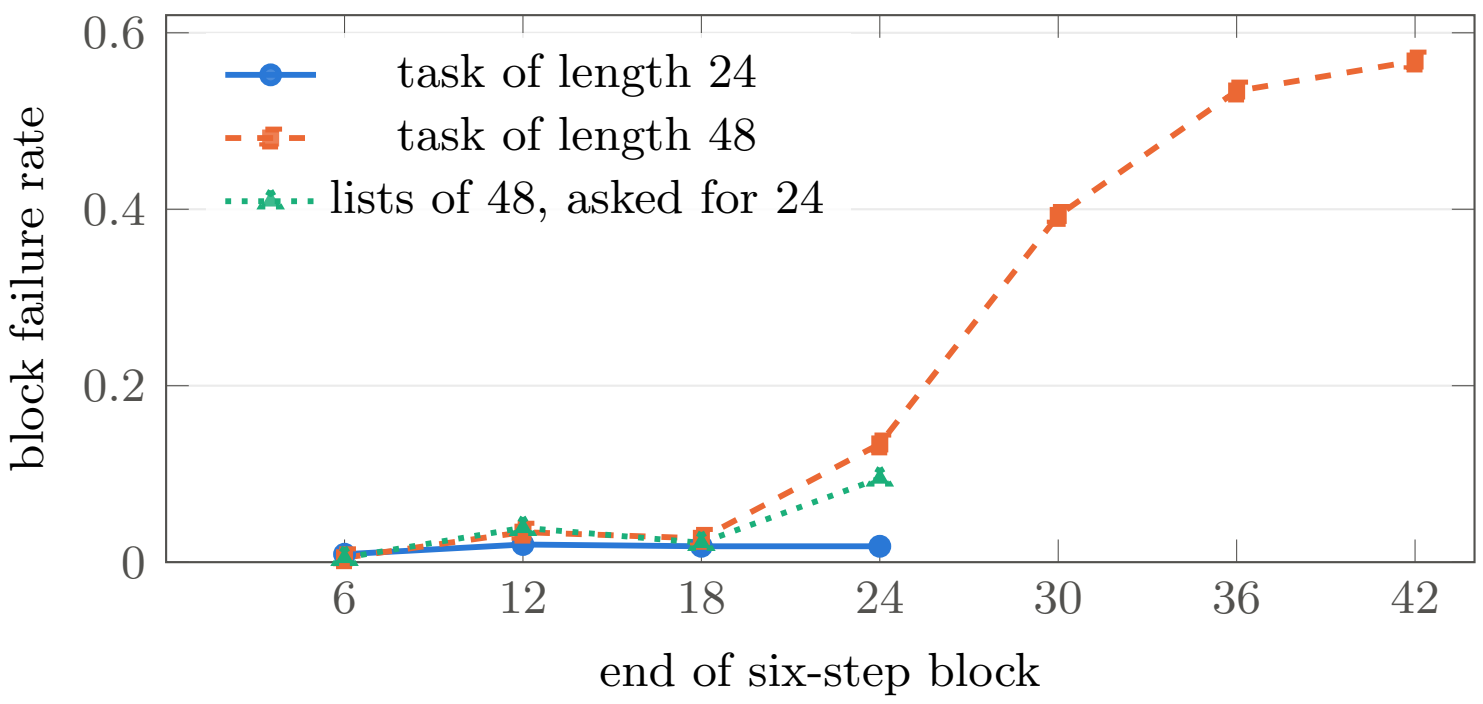
\begin{figure}[t]
\centering
\begin{tikzpicture}
\begin{axis}[width=0.62\linewidth, height=5.2cm, xmin=0, xmax=44, xtick={6,12,18,24,30,36,42}, ymin=0, ymax=0.62,
  xlabel={end of six-step block}, ylabel={block failure rate}, tick label style={font=\footnotesize, color=cInk}, label style={font=\footnotesize},
  axis line style={cInk}, major tick style={cInk}, ymajorgrids, grid style={black!8},
  legend style={font=\footnotesize, at={(0.03,0.97)}, anchor=north west, draw=none, fill=white, fill opacity=0.85, text opacity=1}]
\addplot[cBlue, mark=*, mark size=1.6pt, line width=1.1pt] coordinates {(6,0.009) (12,0.020) (18,0.018) (24,0.018)}; %
\addlegendentry{task of length 24}
\addplot[cOrange, mark=square*, mark size=1.6pt, line width=1.1pt, dashed] coordinates {(6,0.005) (12,0.034) (18,0.027) (24,0.134) (30,0.393) (36,0.534) (42,0.568)}; %
\addlegendentry{task of length 48}
\addplot[cAqua, mark=triangle*, mark size=2pt, line width=1.1pt, dotted] coordinates {(6,0.005) (12,0.039) (18,0.022) (24,0.095)}; %
\addlegendentry{lists of 48, asked for 24}
\end{axis}
\end{tikzpicture}
\caption{S1, recommended decoding, 100 instances, 8 runs each: failure rate within each block of six steps, given the earlier steps correct. The block ending at step 48 is omitted because few runs reach it. On the same steps 19--24, the failure rate depends on how long the task is.}\label{fig:blocks}
\end{figure}

On Qwen3-8B under recommended decoding the violation sits at the end of the shorter task: in S1, $\bar\Delta$ is $-0.003$ at $j=6$ and $+0.020$ at $j=18$, and only reaches $+0.128$ at $j=24$. %
At temperature 1 a small effect is already present at $j=6$, $+0.021$ ($[+0.011,+0.031]$). %
On steps 19--24 the block rate is $0.018$ when these are the last steps of the task, and $0.134$ when the task continues to step 48 (\cref{fig:blocks}). %
The preregistered dose-response check passed at the 5\% level, the second increment only narrowly: $+0.094$ from $24$ to $36$, then $+0.034$ from $36$ to $48$ ($p=0.047$, with a bootstrap interval reaching $-0.003$; \cref{tab:primary}). %

\paragraph{Mostly the length of the input.} S2 splits the effect at $j=24$ into three parts that sum exactly, instance by instance:
\[
\underbrace{+0.095}_{\text{total}} \;=\; \underbrace{-0.033}_{\text{added sentence}} \;+\; \underbrace{+0.115}_{\text{list length }24\to48} \;+\; \underbrace{+0.013}_{\text{requested length }24\to48}. %
\]
Each part is positive when the arm without the change does better, so the negative sign means the point estimate favours the added sentence (\cref{fig:decomp}).

\begin{figure}[ht]
\centering
\begin{tikzpicture}
\begin{axis}[width=0.62\linewidth, height=4.0cm, y dir=reverse, ytick={1,2,3,4}, ymin=0.4, ymax=4.6,
  yticklabels={total, list length, requested length, added sentence}, xmin=-0.09, xmax=0.17,
  xlabel={contribution to $\bar\Delta$ at $j=24$}, xtick={-0.05,0,0.05,0.10,0.15}, xticklabel style={/pgf/number format/fixed, /pgf/number format/precision=2}, tick label style={font=\footnotesize, color=cInk}, label style={font=\footnotesize},
  axis line style={cInk}, major tick style={cInk}, xmajorgrids, grid style={black!8}]
\draw[cInk!60] (axis cs:0,0.4) -- (axis cs:0,4.6);
\addplot[cBlue, line width=1.2pt, forget plot] coordinates {(0.051,1) (0.141,1)}; %
\addplot[cBlue, line width=1.2pt, forget plot] coordinates {(0.079,2) (0.151,2)}; %
\addplot[cBlue, line width=1.2pt, forget plot] coordinates {(-0.035,3) (0.058,3)}; %
\addplot[cBlue, line width=1.2pt, forget plot] coordinates {(-0.062,4) (-0.004,4)}; %
\addplot[cBlue, only marks, mark=*, mark size=2.2pt] coordinates {(0.095,1) (0.115,2) (0.013,3) (-0.033,4)}; %
\end{axis}
\end{tikzpicture}
\caption{S2: the effect of lengthening the task on the first 24 steps, split into three parts that sum exactly to the total, with instance-bootstrap 95\% intervals.}\label{fig:decomp}
\end{figure}
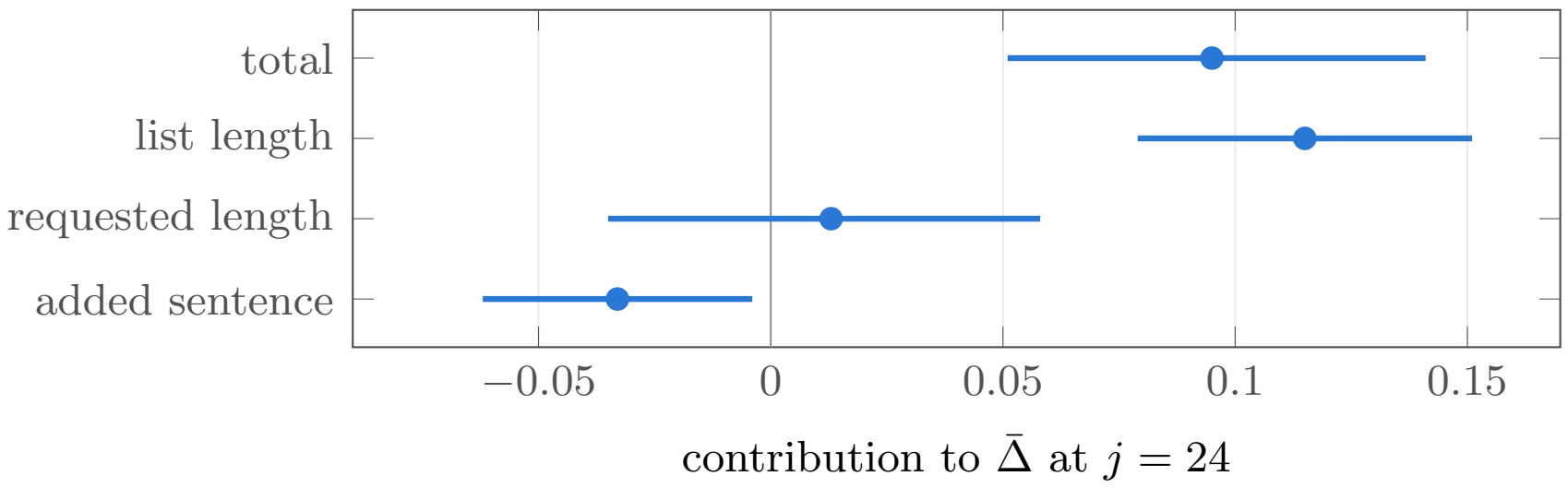
\begin{itemize}\setlength{\itemsep}{0pt}
\item \emph{List length}, with the sentence and the request fixed: $+0.115$ ($[+0.079,+0.151]$), Holm-adjusted $p=0.0001$. %
\item \emph{Requested length}, with lists of length $48$: $+0.013$ ($[-0.035,+0.058]$), Holm-adjusted $p=0.32$. %
\item \emph{The added sentence}: $-0.033$ (Holm-adjusted $p=0.086$). %
\end{itemize}
The requested-length contrast also differs in the added sentence. If the sentence helps equally with long lists (post hoc, untested), the net requested-length effect is $-0.020$ ($[-0.072,+0.035]$). %
S1's preregistered requested-length check had found a small effect, $+0.037$ ($[+0.003,+0.075]$, Holm-adjusted $p=0.030$), with the same sentence in the contrast. %
S2's interval contains that value. On Qwen3-8B we therefore read the two experiments together: most of the violation comes from the longer lists, and the fixed-list contrast, which changes the requested length together with the added sentence, is small. On the second model this last part does not hold (below). %
The list-length contrast adds $144$ prompt tokens, far below the context lengths at which long-context degradation is usually studied \citep{du2025context,liu2024lost}. It also comes with longer outputs: $1.42$ times as many generated tokens. %
The design identifies the total effect of list length. It does not separate retrieving the $k$-th entry from a longer list from producing a longer reasoning trace, and neither does a post hoc classification of the errors (\cref{app:prereg}). What matters for identification is where the effect acts: on steps the model had already been able to do. End-of-task outcomes cannot separate this from a task whose difficulty simply grows with its length (\cref{thm:diag}), and it violates the assumption under which those outcomes are read as within-task hazards.

\paragraph{A second model family.} Experiment~G repeats the paired design on gpt-oss-20b \citep{openai2025gptoss}, under its own preregistration, with the decoding its developers recommend (temperature $1$, top-$p$ $1$), medium reasoning effort, and the same prompts and scoring. A calibration on 20 other instances, with rules fixed in advance, moved the lengths to $48$ and $96$: at length $48$ every calibration run had its first $24$ steps right. On 100 new instances the paired check at $j=48$ rejects, $\bar\Delta=+0.315$ ($[+0.260,+0.367]$; positive for 85 instances, negative for 6), and so does the within-run check (\cref{tab:primary}). %
The same three-part split gives $+0.033$ for the added sentence, $+0.169$ ($[+0.128,+0.210]$, Holm-adjusted $p=0.0001$) for list length and $+0.114$ ($[+0.064,+0.165]$, Holm-adjusted $p=0.0001$) for the fixed-list contrast, which changes the requested length together with the added sentence. %
List length is again the largest part in point estimate, $0.536$ ($[0.395,0.691]$) of the total, but unlike on Qwen3-8B the fixed-list contrast is clearly positive. %
Part of that may come from runs that give no answer at all: at length $96$, 69 of 800 runs gave no parsable answer (21 hit the token budget), against at most 17 in any other arm. They count as failures, as preregistered. In a post hoc worst-case rescoring of all 69 as correct on the first 48 steps, $\bar\Delta$ falls by at most $0.087$, to at least $+0.228$, and the fixed-list contrast to at least $+0.027$. A separate post hoc analysis that drops unparsable runs in every arm and weights instances equally gives $+0.082$ ($[+0.029,+0.134]$) for that contrast and $+0.270$ for $\bar\Delta$. %
On this model the violation is also not confined to the end of the shorter task: on steps 25--30 the block failure rate is $0.072$ at length $96$ and $0.016$ at length $48$. %

\paragraph{Hazard accumulates faster in the second half, for some instances.} The within-run check of \cref{prop:wd} rejects in every experiment: at length $48$, $\hat D(24)=-0.495$ ($[-0.552,-0.438]$) in Pilot~A, $-0.525$ and $-0.632$ in S1, and $-0.604$ in S2 (descriptive there); at length $96$ in G, $\hat D(48)=-0.132$ ($[-0.175,-0.091]$). %
The check needs only independent runs, so these conclusions do not depend on the prefix assumption. By \cref{prop:wd}, $\hat D<0$ says that a positive fraction of instances accumulate more hazard on $(t,2t]$ than on $(0,t]$, not that the hazard rises at every step or for most instances. The check is specific to the length at which it is run: the failure rate at the same steps depends on how long the task is (\cref{fig:blocks}), and by \cref{prop:scale} a change in failure rates alone does not establish a change in shape.

Re-asking steps 25--48 as a separate task of length 24 brings their block rates to $0.014$--$0.046$, the level of the first 24 steps of a task of length 24. %
This is consistent with the list-length effect. The design cannot separate it from removing the model's own reasoning about steps 1--24 from the context. On this task, re-asking the second half from the true $x_{24}$ restores the short-task error rate. In practice this needs a verified intermediate value.

\paragraph{Dependent chains against independent expressions.} S2's preregistered primary check asked whether the length effect is larger on the chain than on independent expressions, which correspond to the composed independent sub-tasks of \citet{li2026fragility}. The independent arm also shows a length effect: $\bar\Delta=+0.036$ ($[+0.015,+0.061]$), with its within-run check also rejecting ($\hat D(24)=-0.346$, $[-0.410,-0.280]$). %
The difference between the arms is positive on the probability scale but negative on the logit scale (\cref{tab:primary}), because the independent arm starts from almost no failures ($0.999$ success at length $24$). %
The preregistered decision table did not anticipate opposite signs on the two scales. We report the comparison as inconclusive, positive on the probability scale and reversed on the logit scale, and make no claim that dependence amplifies the effect. We note a reason found after seeing the data: the two arms also differ in input format, since the independent arm needs no lookup in a list. One preregistered description bears on the mechanism of \citet{li2026fragility}. On the chain, generated tokens per requested step rise from $144$ to $195$ as the length doubles. On independent expressions they fall from $54$ to $48$, in the direction their implicit-budget account predicts; on the chain the effect appears without a fall in this aggregate measure, which counts all generated tokens and does not isolate the thinking budget. %

\paragraph{What end-of-task estimators report on these data.} We fitted the frailty NPMLE of \citet{anon2026pooling} to the data as horizon data would supply them: instances split across lengths, outcome all steps correct. We then compared it with the same estimator applied within runs at length $48$.
\begin{itemize}\setlength{\itemsep}{0pt}
\item In all five comparisons (Pilot A; S1 at both decodings; the chain and the independent arm of S2), the cross-sectional profile interval lies above the within-run interval. For example, in S1 under recommended decoding the cross-sectional estimate is $7.45$ ($[5.85,9.4]$) and the within-run estimate is $4.20$ ($[4.0,4.45]$). %
  The Pilot~A comparison was run after its data had been seen, and it led us to widen the estimator's grid for $k$ (\cref{app:prereg}). The S1 and S2 comparisons were preregistered as descriptive. The preregistration of S1 disclosed a known weakness: in synthetic checks the cross-sectional NPMLE was biased upward and its intervals missed the truth in two of four runs.
\item To check that the direction is not an artefact of the estimator, we ran a post hoc null calibration. We simulated $12$ configurations satisfying the prefix assumption ($k\in\{2,3.5,5\}$, gamma or lognormal frailty, two or four lengths), with $20$ seeds each. The same ordering of intervals appeared in $1$ of $240$ runs. %
\item We do not report the size of the gap. With two lengths the cross-sectional profile often reaches the upper end of the grid for $k$, both on these data and in the null simulations.
\end{itemize}
The preregistered comparison with the shift estimator of \cref{sec:repeated} points the same way. Its cross-sectional estimate of $1/k$ is $0.136$ ($[0.111,0.165]$) in Pilot~A against $0.264$ ($[0.231,0.299]$) within runs. In S1 the corresponding values of $k$ are $5.21$ against $3.58$ at temperature $1$, and $6.30$ against $4.90$ under recommended decoding; the last difference has an interval reaching zero. %
These comparisons are not independent evidence. The paired contrast tests prefix invariance directly; the estimator gaps point the same way, but agreement of the two estimates would additionally require common, correctly specified targets, so the gaps are not separate tests. The null calibration covers only correctly specified Weibull baselines. \Cref{thm:diag} explains why end-of-task data cannot reveal the gap: whenever they are compatible with some prefix-invariant model, they cannot be told apart from the truth, so no estimator built on them alone can recover the within-run shape.

\paragraph{Unpaired data can carry the test.} \Cref{thm:test}(1) needs no pairing. As a post hoc illustration, we split the 100 instances of an experiment in half and observed one half at each length. A two-sample permutation test then rejects at the one-sided 5\% level in $100\%$ of $200$ random splits for S1 and in $76\%$ for S2. %
These are rejection rates over re-splits of one realised data set, not power for new instances. Their point is practical: a benchmark that already repeats runs needs only to record whether a run is still correct at a fixed checkpoint, and to generate tasks with comparable prefixes. Pairing adds the instance-level hypothesis and the exact test of \cref{thm:test}(3)--(4).

\paragraph{Limits.}
\begin{itemize}\setlength{\itemsep}{0pt}
\item Two open models from different families and one task family, and the tasks are not agentic; the comparisons with end-of-task estimators use the first model only. A Tower of Hanoi pilot, in which instances differed only by relabelling and failures concentrated on one systematic error, was not used for the prefix test.
\item Carrying the conclusion to horizon data needs a bridge from ``asked for $c$ steps'' to a task's rated human time. Whether agent tasks of different rated length have comparable prefixes is itself an empirical question.
\item The two independent sets of instances under the same condition differ in level. On steps 19--24 of a task of length 24, the block rate is $0.018$ in S1 and $0.054$ in S2. %
\item Analyses not fixed in the preregistrations are labelled as post hoc where they appear.
\end{itemize}

\section{An application: METR's run-level data, read through the diagonal}\label{sec:metr}

This section applies the estimator of \cref{sec:repeated} to public horizon data. By \cref{thm:diag,cor:stat} what it measures is the shape of the diagonal, the end-of-task success curve; it is a within-task shape only under the prefix assumption, which \cref{sec:experiments} found to fail on a controlled task, in the direction of the length-effect term of the shared-frailty decomposition in \cref{rem:frailty}. The results are reported with that reading.

\paragraph{What can bias the estimate.} Five mechanisms can bias the repeated-run estimator: correlated runs, selection on the number of runs, error in human times, outcome-based task filtering, and task composition. \Cref{app:mechanisms} states them as lemmas and records the direction each one moves the estimate, where a direction can be given, and the data feature that removes it (\cref{tab:mech}). The reanalysis below uses these results.

\paragraph{Data and pipeline.} We use the two public run-level releases (TH1.0 and TH1.1) of METR's time-horizon data \citep{kwa2025measuring}, with 170 and 228 tasks and 41{,}629 and 24{,}008 distinct run identifiers; 155 tasks and 1{,}707 runs are shared, so the releases are not independent replications. Runs flagged with fatal errors are excluded (one specification keeps them). For each model we use tasks with $t_i>0$ and $n_i\ge2$ and keep models with at least 60 such tasks: 20 models in TH1.1 (193--228 tasks each) and 33 in TH1.0 (66--169). Weights are $w_i=1/\sqrt{\text{number of tasks in the family of }i}$, counted within the model's analysis set, and 95\% percentile intervals come from 1{,}000 bootstrap resamples of task families. The raw releases contain 79 (TH1.1) and 50 (TH1.0) families; a model's base analysis set has 63--79 families in TH1.1 and 5--49 in TH1.0 (medians 78 and 41), and the number changes with the specification. Every point estimate and every bootstrap draw is checked by a status gate (no separation, an interior second moment, converged and ridge-stable fits, a negative slope); an interval with more than 5\% failed draws is marked and excluded from all counts. Frontier models are those with a base 50\% horizon of at least 100 minutes (10 in TH1.1, 5 in TH1.0). Details are in \cref{app:pipeline}.

\paragraph{Assumptions.} All results below are conditional on: model $\cM$ with a log-logistic cross-section, including a distribution of task difficulty that does not change with task length ($Z\indep t$)---for statements about the direction of the hazard, $\cM$ can be replaced by the general frailty model of \cref{prop:direction}, but the log-logistic cross-section stays; scheduling, reruns and fatal-error exclusion non-informative about outcomes; a factorising selection probability for each analysis set defined by the number of runs (\cref{lem:cross}); task composition not breaking (T) (\cref{lem:comp}). Results from \emph{all pairs} of runs require full A0; results from \emph{cross-batch pairs} require only the weaker conditions of \cref{lem:cross}. Batches are formed by single-linkage clustering of start times with a one-hour gap. None of these assumptions is verified by the data; the status gate is an engineering check, and passing it does not establish them. Intervals are per model and per specification, without correction for multiplicity.

\paragraph{Correlated runs.} Within tasks that have both same-batch and cross-batch pairs, let $D_i=\hmtwo_{2,\mathrm{same},i}-\hmtwo_{2,\mathrm{cross},i}$ be the difference in the rate of double success between the two kinds of pair. The normalised difference $\hat\rho_{\mathrm{ex}}=\sum_iw_iD_i/\sum_iw_i(\hat p_i-\hmtwo_{2,\mathrm{cross},i})$ estimates the excess within-batch correlation when cross-batch runs are conditionally independent. Pooled over all models it is $-0.037$ $[-0.182, 0.087]$ in TH1.1 and $0.008$ $[-0.111, 0.120]$ in TH1.0; for the frontier groups it is $-0.114$ $[-0.349, 0.074]$ and $0.166$ $[-0.427, 0.819]$. The residual correlation that would by itself move a frontier model's estimate to $1/k=1$ ranges over $[0.256, 0.649]$ in TH1.1 and $[0.567, 0.770]$ in TH1.0. In TH1.1 the upper bound for $\rho_{\mathrm{ex}}$ lies below the smallest such value; in TH1.0 it does not, so for the TH1.0 frontier group the all-pairs results rest on an untested part of A0, and we rely on cross-batch pairs. The test has low power: in simulations on the observed designs it rejects at rates $0.05$--$0.13$ when the true excess correlation is $0.1$.

\paragraph{Result 1: no evidence of declining within-task hazard (under the prefix assumption).} The formal evidence consists of the all-pairs specifications other than the two outcome-based filters (252 cells in TH1.1 and 374 in TH1.0 that pass the gate without the failed-draw mark), percentile and BCa intervals for the base specification, and the cross-batch and cross-day subsamples ($8+3=11$ rows in TH1.1, $20+12=32$ in TH1.0). In none of them does the interval for $1/k$ lie entirely above one. This is a failure to detect $k<1$, not evidence against it. Three further analyses, reported as descriptive and not as support, show the same: the two outcome-based filters (40 and 64 cells), a quadratic-logit cross-section (exploratory), and a nonparametric monotone cross-section whose intervals all contain one and carry little information.

\begin{figure}[t]
\centering
\begin{tikzpicture}
\begin{groupplot}[
  group style={group size=2 by 1, horizontal sep=1.6cm},
  width=0.49\linewidth, height=6.2cm,
  xmode=log, xmin=0.1, xmax=2000,
  ymin=0, ymax=2.3,
  xlabel={base 50\% horizon $\hat h$ (minutes)},
  tick label style={font=\footnotesize, color=cInk}, label style={font=\footnotesize},
  title style={font=\small},
  axis line style={cInk}, major tick style={cInk},
  ymajorgrids, grid style={black!8},
  unbounded coords=discard,
  error bars/error bar style={line width=0.5pt, opacity=0.7},
  error bars/error mark options={draw=none},
]
\nextgroupplot[title={TH1.1 (20 models)}, ylabel={$1/\hat k$ and 95\% interval}]
\addplot[cInk, dashed, forget plot] coordinates {(0.1,1) (2000,1)};
\addplot[cBlue, only marks, mark=*, mark size=1.6pt,
  error bars/.cd, y dir=both, y explicit]
  table[col sep=comma, comment chars={\#}, x=h50, y=base_est,
        y error plus expr=\thisrow{base_hi}-\thisrow{base_est},
        y error minus expr=\thisrow{base_est}-\thisrow{base_lo}] {figdata/metr_TH1.1.csv};
\addplot[cOrange, only marks, mark=triangle*, mark size=2.0pt,
  error bars/.cd, y dir=both, y explicit]
  table[col sep=comma, comment chars={\#}, x expr=\thisrow{h50}*1.35, y=xb_est,
        y error plus expr=\thisrow{xb_hi}-\thisrow{xb_est},
        y error minus expr=\thisrow{xb_est}-\thisrow{xb_lo}] {figdata/metr_TH1.1.csv};
\draw[cInk, densely dotted] (axis cs:100,0) -- (axis cs:100,2.3);
\node[font=\scriptsize, color=cInk, anchor=north east] at (axis cs:95,2.25) {frontier $\to$};
\nextgroupplot[title={TH1.0 (33 models, 32 eligible)},
  legend style={font=\footnotesize, at={(0.03,0.97)}, anchor=north west, draw=none, fill=white, fill opacity=0.85, text opacity=1}]
\addplot[cInk, dashed, forget plot] coordinates {(0.1,1) (2000,1)};
\addplot[cBlue, only marks, mark=*, mark size=1.6pt,
  error bars/.cd, y dir=both, y explicit]
  table[col sep=comma, comment chars={\#}, x=h50, y=base_est,
        y error plus expr=\thisrow{base_hi}-\thisrow{base_est},
        y error minus expr=\thisrow{base_est}-\thisrow{base_lo}] {figdata/metr_TH1.0.csv};
\addlegendentry{all pairs (base)}
\addplot[cOrange, only marks, mark=triangle*, mark size=2.0pt,
  error bars/.cd, y dir=both, y explicit]
  table[col sep=comma, comment chars={\#}, x expr=\thisrow{h50}*1.35, y=xb_est,
        y error plus expr=\thisrow{xb_hi}-\thisrow{xb_est},
        y error minus expr=\thisrow{xb_est}-\thisrow{xb_lo}] {figdata/metr_TH1.0.csv};
\addlegendentry{cross-batch pairs}
\draw[cInk, densely dotted] (axis cs:100,0) -- (axis cs:100,2.3);
\end{groupplot}
\end{tikzpicture}
\caption{Estimated $1/k$ against the base 50\% horizon for every model with an eligible base interval (all 20 in TH1.1; 32 of 33 in TH1.0, GPT-2 being marked). Circles: all pairs of runs, base specification. Triangles (shifted right for legibility): pairs of runs in different batches, shown where that subsample passes the status gate without the failed-draw mark. Dashed line: constant within-task hazard; below it, $k>1$. Intervals are per model, without multiplicity correction. The decline of $1/\hat k$ with $\hat h$ is not interpreted: effects of where the design sits on the curve have not been excluded. Plotted values are produced by \texttt{s3\_paper\_numbers.py} (\cref{app:prov}).}\label{fig:metr}
\end{figure}
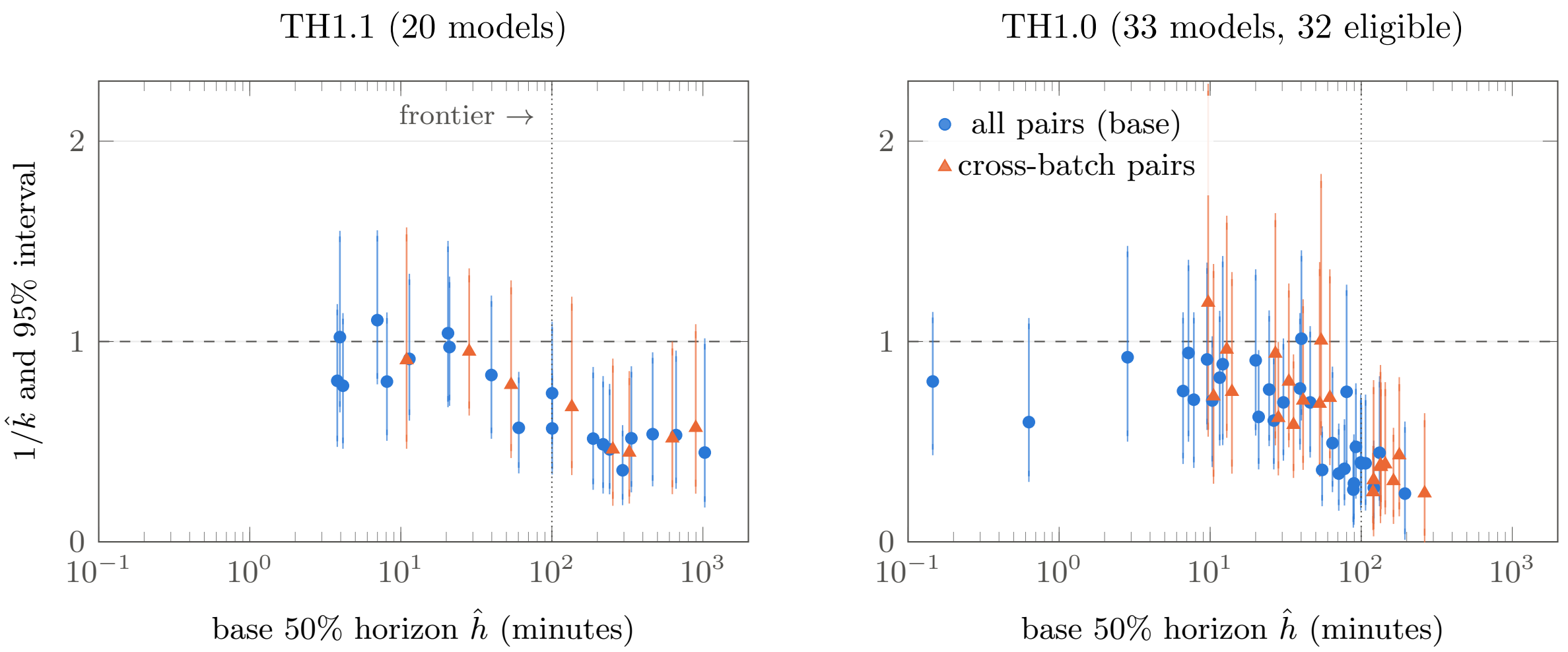

\paragraph{Result 2: frontier models, $k\le1$ rejected in a limited range (under the prefix assumption).} \Cref{tab:counts} counts frontier models whose interval lies entirely below one; \cref{fig:metr} shows all eligible models. In TH1.0, all five frontier models have cross-batch intervals below one (\cref{tab:frontier} in \cref{app:pipeline}), and so do their all-pairs intervals in eight of the twelve estimable specifications. In TH1.1 the cross-batch analysis is possible for only five frontier models, of which three have intervals below one; with baseline-timed tasks only, no frontier interval lies below one. Over all eligible TH1.1 models (20, 20 and 13 in the three specifications, so not the same set), median $1/\hat k$ is $0.66$ in the base specification, $0.78$ with baseline-timed tasks and $0.98$ with estimated times; composition (\cref{lem:comp}) and measurement error (\cref{lem:me}) are candidate explanations, and neither has been identified.

Part of the TH1.0 result may already be in the cross-section (we report no interval for $\beta$). The point estimate $\hat\beta$ exceeds one for 15 of the 32 eligible TH1.0 models (\cref{app:prov}), including the frontier models o3 ($1.068$) and GPT-5 ($1.110$), and within $\cM$ a log-logistic cross-section with $\beta>1$ implies $k>1$ (\cref{prop:kbound}); without the Weibull form it excludes a concave baseline (\cref{prop:direction}). Repeated runs add most for the three other TH1.0 frontier models, whose $1/\hat\beta$ exceeds one (\cref{tab:frontier}), and for TH1.1, where only one of the ten frontier models (\cref{tab:frontier}) and one of the 20 eligible models (\cref{app:prov}) has $\hat\beta>1$. By \cref{prop:direction}, for every frontier model whose cross-batch interval lies below one (all five in TH1.0, three of five in TH1.1), that interval rejects a non-increasing average hazard without the Weibull form, given a log-logistic cross-section; it is a statement about a within-task hazard only under the prefix assumption.

\begin{table}[t]
\centering
\small
\caption{Frontier models whose 95\% interval for $1/k$ lies entirely below one, over frontier models estimable without the failed-draw mark.}\label{tab:counts}
\begin{tabular}{@{}llcc@{}}
\toprule
Specification & Pairs & TH1.0 & TH1.1 \\
\midrule
Base & all & 5/5 & 9/10 \\
Cross-batch & cross-batch & \textbf{5/5} & 3/5 \\
No SWAA / HCAST only / METR weights & all & 5/5 each & 9/10 each \\
$t\ge2$ min / version-stable / no cloned tasks & all & 5/5 each & 8/10 each \\
Equal weights & all & 5/5 & 6/10 \\
$t\ge8$ min / fatal runs kept & all & 4/5 each & 9/10 each \\
$t\le240$ min & all & 3/5 & 8/9 \\
Baseline-timed tasks only & all & 3/5 & \textbf{0/10} \\
Estimated times only & all & -- & 0/6 \\
\bottomrule
\end{tabular}
\end{table}

\paragraph{Specification checks.} The third moment gives a second shift, $m_3(x)=S(x+\theta\log_23)$. The two estimates differ significantly for 5 of 20 models in TH1.1 (one of ten frontier models) and 3 of 31 in TH1.0 (none of five frontier models). \Cref{cor:logit} also requires $1/\hat k\le1/\hat\beta$; among the 52 eligible models only davinci-002 violates it in point estimate ($0.801>0.693$). A quadratic-logit cross-section is explored but not used as evidence. Filtering out tasks that all or no models solve is reported for comparison only: it is outcome-based selection (see the paragraph after \cref{lem:cross}), and in simulations on the observed designs the bias of the filtered estimate relative to the true $\theta$ is $+0.149$ to $+0.322$, against $-0.021$ to $+0.010$ without filtering (\cref{app:pipeline}).

\paragraph{Error in human times (exploratory).} With Barry's noise model for human times \citep{barry2026modelling} and linear SIMEX extrapolation, the median $1/\hat k$ of frontier models moves from $0.516$ to $0.430$ in TH1.1 and from $0.392$ to $0.343$ in TH1.0; the extrapolated value is below the observed one for every frontier model. Under classical error this is the direction in which Result 2 is conservative. We report no intervals, and by \cref{lem:me} the same pattern would appear under Berkson error, where it would be an over-correction.

\section{Discussion and limitations}\label{sec:discussion}

\paragraph{What $k>1$ means.} Two comparisons should be kept apart. In $\cM$ with a log-logistic cross-section of slope $\beta$, $k\ge\beta$ always holds, and the task-level complementary log-log slope $k$ (in $\log t$) exceeds the marginal one, $\beta u/\{(1+u)\log(1+u)\}$ with $u=(t/h)^\beta$, which is below $\beta$ for every $t$, even at $k=\beta$. What distinguishes $k>\beta$ from $k=\beta$ is the frailty: for a log-logistic cross-section, $k=\beta$ needs an exponential frailty and $k>\beta$ one with infinite mean (\cref{prop:kbound}). Neither says anything about the direction of the hazard; $\beta<k<1$ is a declining hazard. The statement about direction is $k>1$, which within $\cM$ and under the assumptions of \cref{sec:metr} is identified from repeated runs. Read with the prefix assumption, $k>1$ is a hazard that rises within a task, as with compounding errors. Read without it, $k$ is only the slope of a task's complementary log-log failure probability in $\log t$ given its frailty, and $k>1$ says that, at fixed frailty, the negative log success probability of a task grows faster than proportionally to human time---for instance because the difficulty a task poses to the agent grows faster than proportionally with human time. METR-type data are consistent with both readings, and by \cref{thm:diag,prop:compat} no estimator built on end-of-task data can choose between them while the data remain compatible with a prefix-invariant model. On a controlled task (\cref{sec:experiments}) the second reading is not hypothetical: on average, the same instance fails more often on steps it shares with a shorter task, largely because the lists it must index are longer (about half of the effect on the second model), and, on the first model, end-of-task estimators report a steeper shape than the within-run one. Whether agent tasks that take humans longer also present longer or more demanding inputs at their start is an empirical question about agent benchmarks. Carrying the conclusion to horizon data needs a bridge from ``asked for $c$ steps'' to rated human time. What the present data do settle is mostly negative: the cross-section cannot show that hazards decline within tasks (\cref{prop:kbound}), and the repeated-run evidence, under its assumptions, does not show it either. For the five TH1.0 frontier models it goes further: their cross-batch intervals lie entirely below one, each interval taken on its own and without multiplicity correction. Under the assumptions of \cref{sec:metr} and the prefix assumption this is evidence against a non-increasing average within-task hazard for these models, but only if the prefix assumption holds, and \cref{sec:experiments} shows that it can fail. By \cref{prop:direction} the conclusion does not need the Weibull form, but it still needs the log-logistic cross-section (limitation (iii)); for o3 and GPT-5 the cross-section already points the same way. The direction of the violation found in \cref{sec:experiments}, harder common steps in longer tasks, is the one that makes end-of-task data report a larger $k$ (\cref{rem:frailty}, under a shared frailty).

\paragraph{For evaluation designers.} Benchmarks that already repeat runs can make the prefix assumption testable at little cost. They need to record whether each run is still correct at one or more checkpoints fixed in advance, and to generate tasks of different lengths whose prefixes are comparable by design (\cref{thm:test}). Running some instances at two lengths gives an exact paired test. On our task, re-asking the second half as a new task from the true intermediate value restored the short-task error rate. In practice this needs a verified intermediate value.

\paragraph{Limitations.}
(i) A0 is not verified; the excess-correlation test has low power and fails its threshold for the TH1.0 frontier group, and cross-batch pairs do not protect against correlation shared across batches, outcome-dependent reruns or non-factorising selection. Residual correlation of the size of the values above would overturn the frontier results.
(ii) The frontier result is most robust in the TH1.0 cross-batch analysis, which was not stratified by the source of human times; for Claude Opus 4.5 and GPT-5.1-Codex-Max every pair of runs on the tasks entering it is a cross-batch pair, so there the two estimates, computed on that same restricted subsample, coincide. TH1.1 depends on the source of human times, and only five TH1.1 frontier models support a cross-batch estimate.
(iii) \Cref{prop:direction} removes the Weibull form but not the log-logistic cross-section used by the estimator; the nonparametric cross-section we tried is uninformative, and a nonparametric one-sided test of $\theta(x)\ge1$ is left open.
(iv) Tasks with both kinds of pairs come only from task suites with time stamps (HCAST, RE-Bench).
(v) Many intervals are reported without multiplicity correction; repeated zeros across specifications are not independent pieces of evidence.
(vi) Our pipeline differs from METR's official one and does not apply SIMEX.\\ (vii) The experiments use two open models and one task family, which is not agentic (\cref{sec:experiments}).

\section*{Use of AI tools}
AI assistants took part in discussing the choice of research direction; the author decided the main line of the work and carried it forward. AI assistants, mainly Claude (Anthropic), with DeepSeek and Codex (OpenAI) for review, carried out much of the mathematical derivation, wrote and checked the code, drafted and revised the text, and reviewed the proofs and analyses. The author has personally checked the proofs and results and takes full responsibility for the content.

\section*{Code and data availability}
{\raggedright All empirical results are produced by scripts available at \url{https://github.com/Zrq23/within-task-hazards}; \cref{app:prov} states where each kind of number comes from.\par}

\renewcommand{\bibfont}{\small}
\setlength{\bibsep}{2pt}
\bibliographystyle{plainnat}
\bibliography{refs}

\clearpage
\appendix
\crefalias{section}{appendix}

\section{What moves the shift}\label{app:mechanisms}\label{sec:mechanisms}

\Cref{prop:repr} reduces every question about the bias of (E) to the sign and size of $G(\theta)$. \Cref{tab:mech} collects five mechanisms; three are stated as lemmas here and task composition in \cref{app:proofs}. All directions refer to population limits; sampling error is handled separately by the bootstrap of \cref{sec:metr}.

\begin{lemma}[Correlated runs]\label[lemma]{lem:a0}
Let run $a$ of task $i$ have conditional success probability $p_{ia}$ with mean $\bar p_i$ over runs, let $\bar C_i$ be the average conditional covariance over ordered pairs of distinct runs, and put $\bar c_i=\bar C_i-\Var_{\mathrm{pop}}(p_{ia})/(n_i-1)$ with $\Var_{\mathrm{pop}}(p_{ia})=n_i^{-1}\sum_a(p_{ia}-\bar p_i)^2$. Then $\E\hat p_i=\bar p_i$ and $\E\hmtwo_{2,i}=\bar p_i^2+\bar c_i$. The first step does not depend on $\bar c$, $\hat\delta$ is a strictly increasing function of $\Delta_w=\sum_iw_i\bar c_i$, and, since $\hat\gamma<0$, $\htheta$ is strictly decreasing in $\Delta_w$. Hence $\Delta_w>0$---for example positive residual correlation between runs whose success probability does not drift---makes $1/\hat k$ too small; drift across runs enters $\bar c_i$ with a negative sign.
\end{lemma}

The proof expands $\E\,y_i(y_i-1)$ (\cref{app:proofs}). Correlation within a batch of runs launched together---shared infrastructure, near-deterministic decoding---is the most plausible violation of A0 in agent evaluations, and without drift it biases the estimate towards \emph{finding} an increasing hazard. Runs in different batches avoid it:

\begin{lemma}[Cross-batch pairs and selection on runs]\label[lemma]{lem:cross}
Suppose that given $p_i$, runs in different batches are independent (within a batch any dependence is allowed), batch labels are non-informative about outcomes, and every batch has conditional mean $p_i$. Then the average $\hmtwo_{2,\mathrm{cross},i}$ of $Y_{ia}Y_{ib}$ over pairs in different batches satisfies $\E[\hmtwo_{2,\mathrm{cross},i}\mid p_i,\text{selected}]=p_i^2$. If, moreover, the marginal probability that a task enters the analysis set, $\bar\psi(z,t)=\sum_n\Pr(n\mid z,t)\,\psi(z,t,n)$, factorises as $\psi_t(t)\psi_z(z)$, the selected tasks again follow $\cM$ with the same $k$, and (T) holds on the analysis set. Factorisation is sufficient but not necessary.
\end{lemma}

The selection condition applies to every analysis set defined by the number of runs, not only to the cross-batch subsample: tasks with $n_i\ge2$ enter every analysis, and the third-moment check uses $n_i\ge3$ (169--228 tasks per model in TH1.1 and 66--169 in TH1.0; in TH1.0 the range happens to coincide with that for $n_i\ge2$). If, say, harder long tasks are rerun more often, (T) can fail on the analysis set: with $\psi=1-\exp\{-ZH_0(t)\}$ the limit of $\htheta$ is $1.1436$ instead of $1$, and with $\psi=\exp\{-ZH_0(t)\}$ it is $1.7305$. A selection that is not separable but periodic in $x$ with period $\theta$ preserves (T), so factorisation is not necessary. The same argument covers retention that depends on the outcomes of other models: a factorising retention probability $\chi(z,t)$ preserves (T), while a non-factorising one, such as dropping tasks that all or no other models solve, generally breaks it---selection bias in the sense of \citet{berkson1946limitations} and \citet{hernan2004structural}. None of these conditions can be checked from outcomes alone.

\begin{lemma}[Error in human times]\label[lemma]{lem:me}
Let $X=\log_2t$ be the true and $W$ the recorded log time, with non-differential error, let (T) hold on the true scale and put $A(\omega)=\E[S(X)\mid W=\omega]$, $B(\omega)=\E[m_2(X)\mid W=\omega]$. If $X\mid W=\omega$ is distributed as $a+b\omega+U$ with $b>0$ and $U\indep W$, then $B(\omega)=A(\omega+\theta/b)$. For classical error $W=X+\varepsilon$ with $X\sim N(\mu,\tau^2)$ and $\varepsilon\sim N(0,\sigma^2)$ independent, $b=\tau^2/(\tau^2+\sigma^2)<1$ and the observed shift is inflated to $\theta/b$; for Berkson error $X=W+\varepsilon$ with $\varepsilon\indep W$, $b=1$ and the shift is unchanged \citep{berkson1950two,carroll2006measurement}.
\end{lemma}

Up to the projection residual of \cref{prop:repr}, classical error therefore makes $1/\hat k$ too large, that is, it works against finding an increasing hazard. The error variance $\sigma^2$ is not identified from $(W,Y)$ in this nonparametric setting and must come from outside (see \citealp{gustafson2003measurement} for adjustment using external or prior information on the error); we therefore use the lemma for a sensitivity analysis rather than a correction. It also shows a limit of SIMEX \citep{cook1994simulation}: adding noise makes the SIMEX curve rise under Berkson error as well, because $X-W'$ is then correlated with the noisier $W'$, and extrapolating to $\zeta=-1$ over-corrects (by $5.5\%$ and $9.9\%$ at $\sigma=1.13$ and $1.51$ in a homoscedastic Gaussian example; \cref{tab:mech} instead reports the classical-error bias without extrapolation). A rising SIMEX curve does not distinguish classical from Berkson error.

\begin{table}[t]
\centering
\small
\caption{Mechanisms that move the population limit of $\htheta=1/\hat k$. ``Example'' gives a deterministic toy computation (true value in parentheses where the entry is a limit of $\htheta$); sources in \cref{app:prov}.}\label{tab:mech}
\begin{tabularx}{\linewidth}{@{}>{\raggedright\arraybackslash}p{2.7cm}>{\raggedright\arraybackslash}X>{\raggedright\arraybackslash}p{2.9cm}>{\raggedright\arraybackslash}p{3.4cm}@{}}
\toprule
Mechanism & Effect on the shift & Example & What the data can do \\
\midrule
Correlated runs (\cref{lem:a0}) & $\htheta$ strictly decreasing in $\Delta_w$; $\Delta_w>0$ gives too small $1/\hat k$ & $0.373$ / $0.628$ ($0.5$)\textsuperscript{a} & cross-batch pairs remove within-batch correlation only \\
Selection on number of runs (\cref{lem:cross}) & (T) kept if $\bar\psi$ factorises; otherwise it can fail, in a direction the lemma does not determine (both examples push $\htheta$ up) & $1.1436$ / $1.7305$ ($1$) & none; can compare selected and unselected tasks \\
Task composition changing with length (\cref{lem:comp}) & (T) kept under the conditions of note b; either direction otherwise & $0.7045$ / $0.3239$ ($0.5$) & stratify; changes the weighting \\
Error in human times (\cref{lem:me}) & classical: shift inflated to $\theta/b$; Berkson: unchanged & $+5.8\%$ / $+10.9\%$\textsuperscript{c} & sensitivity analysis with external $\sigma^2$ \\
Outcome-based task filtering (\cref{lem:cross}) & non-factorising retention generally breaks (T); factorising retention keeps it & $1.6869$ ($1$)\textsuperscript{d} & do not filter on outcomes \\
\bottomrule
\end{tabularx}
\par\smallskip
{\footnotesize\raggedright
\textsuperscript{a}\,Constant correlation $\pm0.1$.\quad
\textsuperscript{b}\,With a common $\theta$ across sources, (T) is kept if source shares do not vary with length or sources share one curve.\quad
\textsuperscript{c}\,Classical error, no extrapolation, at $\sigma=1.13$ / $1.51$.\quad
\textsuperscript{d}\,Dropping tasks all/no models solve.\par}
\end{table}

\section{Proofs and further statements}\label{app:proofs}

\paragraph{Proof of \cref{prop:cm}.} $f(u)=1/(1+u)$ and $g(u)=e^{-u}$ are completely monotone on $(0,\infty)$. For $0<\beta\le1$, $\phi(t)=(t/h)^\beta$ is a Bernstein function: $\phi\ge0$ and $\phi'(t)=\beta h^{-\beta}t^{\beta-1}$ is completely monotone. A completely monotone function composed with a Bernstein function is completely monotone, so $S=f\circ\phi$ is, and Bernstein's theorem gives the mixture (both results are in \citealp{schilling2012bernstein}); $S(0+)=1$ and $S(\infty)=0$ make $\mu$ a probability measure on $(0,\infty)$. A mixture of exponentials is log-convex (by the Cauchy--Schwarz inequality applied to its mixture representation), so its hazard $-(\log S)'$ is non-increasing, whereas for $\beta>1$ the log-logistic hazard $\beta t^{\beta-1}h^{-\beta}/\{1+(t/h)^\beta\}$ increases near $0$. The Weibull case is identical with $g$ in place of $f$.

\paragraph{Proof of \cref{prop:kbound}.} We need $L_Z(u)=S(H_0^{-1}(u))=1/(1+cu^{\alpha})$ with $c=(s/h)^\beta$ and $\alpha=\beta/k$. If $\alpha\le1$, take $Z=c^{1/\alpha}X$ with $X$ Mittag-Leffler of index $\alpha\in(0,1]$ in the sense of \citet{pillai1990mittag}, whose Laplace transform is $1/(1+u^\alpha)$ (later also called the positive Linnik law; it is geometrically stable but is not the positive stable law; exponential when $\alpha=1$). If $\alpha>1$, $-L_Z'(0+)=\lim_{u\downarrow0}\alpha cu^{\alpha-1}/(1+cu^\alpha)^2=0$, and by monotone convergence $-L_Z'(0+)=\E Z$, so $Z=0$ almost surely and $L_Z\equiv1$, a contradiction. The Weibull case uses $L_Z(u)=\exp(-cu^{\kappa/k})$, the Laplace transform of a scaled positive stable law of index $\kappa/k$ when $\kappa\le k$ \citep{hougaard1986survival}, and the same mean argument when $\kappa>k$.

\begin{remark}[Identifying $H_0$; frailty versions of known copula results]\label[remark]{rem:three}
Let $p_i=\exp\{-Z_iH_0(t_i)\}$ with $H_0$ continuous and strictly increasing, $H_0(0+)=0$, $H_0(\infty)=\infty$, $Z\indep t$ and $\Pr(Z>0)>0$, and suppose the population curves $S$ and $m_r$ are known on $(0,\infty)$. Then $\psi_r:=S^{-1}\circ m_r$ satisfies $H_0(\psi_r(t))=rH_0(t)$. If $(\tilde Z,\tilde H_0)$ is another such pair producing the same curves, $K:=\tilde H_0\circ H_0^{-1}$ is continuous, strictly increasing and satisfies $K(ru)=rK(u)$, and $L_{\tilde Z}=L_Z\circ K^{-1}$.
(a) With $r=2$ and $r=3$: $u\mapsto\log_2K(u)-\log_2u$ is invariant under $u\mapsto2u$ and $u\mapsto3u$; since $\log_23$ is irrational these generate a dense group of dilations, so by continuity $K$ is linear and $H_0$ is identified up to a multiplicative constant, which can be absorbed into $Z$. In copula terms this is the result that the diagonal of a three-dimensional Archimedean copula and that of its two-dimensional margin determine the generator \citep[as reported in Remark~18 of \citealp{foschi2021diagonal}]{jaworski2009copulas}.
(b) With $r=2$ alone the functional equation allows $K(u)=u\,g(\log_2u)$ with $g$ positive and $1$-periodic, and strictly increasing non-linear solutions exist (for example $g(v)=1+0.01\sin(2\pi v)$). If $\E Z<\infty$ they are excluded. Put $q(v)=(1-L_{\tilde Z}(v))/v=\E[(1-e^{-v\tilde Z})/v]$; since $(1-e^{-vz})/v$ is non-increasing in $v$ for every $z\ge0$, $q(v)$ increases to $\E\tilde Z\in(0,\infty]$ as $v\downarrow0$, so it has a unique limit in the extended reals. Since $g$ is continuous, positive and periodic, it is bounded away from $0$ and $\infty$. For any phase $a$, put $w_n=2^{a-n}$; then $K(w_n)=w_ng(a)\downarrow0$ and $q(K(w_n))=\{(1-L_Z(w_n))/w_n\}/g(a)\to\E Z/g(a)$, which is finite and positive. Uniqueness of the limit forces $g(a)$ to be the same for every $a$, so $K$ is linear. In copula terms, with generator $\phi=L_Z^{-1}$ and diagonal $\delta(u)=L_Z(2L_Z^{-1}(u))$, $\E Z<\infty$ is equivalent to $\lim_{t\uparrow1}\phi'(t)<0$, the derivative condition in Theorem~2.3 of \citet{schwarz2013identifiability}, and it implies the weaker condition $\delta'(1^-)=2$ under which the diagonal determines the generator \citep[as reported in Remark~18 of \citealp{foschi2021diagonal}]{erdely2014frank}. A frailty with density $z^{-2}$ on $[1,\infty)$ has $\E Z=\infty$ and still $\delta'(1^-)=2$, so $\E Z<\infty$ is sufficient but not necessary for the diagonal to determine the generator; in the case of (c) below, $\delta'(1^-)=2^\alpha<2$.
(c) Without $\E Z<\infty$, $S$ and $m_2$ need not identify $H_0$; this happens for the frailties of \cref{prop:kbound} with $k>\beta$. Take $L_Z(u)=1/(1+cu^\alpha)$ with $0<\alpha<1$, the log-logistic case with $k>\beta$, and the Bernstein function $\varphi(v)=\sum_{n\in\mathbb Z}2^{-n\alpha}(1-e^{-v2^n})$, for which $\varphi(2v)=2^\alpha\varphi(v)$. Then $K^{-1}(v):=(\varphi(v)/c)^{1/\alpha}$ is continuous and strictly increasing from $0$ to $\infty$ and satisfies $K^{-1}(2v)=2K^{-1}(v)$, and $L_Z\circ K^{-1}=1/(1+\varphi)$ is completely monotone, so it is the Laplace transform of a frailty $\tilde Z$; the pair $(\tilde Z,K\circ H_0)$ produces the same $S$ and $m_2$. $K$ is linear only if $F(y)=\varphi(2^y)2^{-\alpha y}$ is constant, but $F$ is $1$-periodic and, by Poisson summation and $\int_0^\infty(1-e^{-u})u^{-s-1}du=-\Gamma(-s)$ for $0<s<1$, has Fourier coefficients $-\Gamma(-\alpha-2\pi im/\ln2)/\ln2$, which are non-zero because $\Gamma$ has no zeros. The same construction applies to the Weibull case with $e^{-\varphi}$. The perturbation is tiny: the relative amplitude of $F$ is about $2|\Gamma(-\alpha-2\pi i/\ln2)|/|\Gamma(-\alpha)|$, close to $10^{-7}$ for $\alpha=0.3$ and $0.5$. And the doubling map of \cref{prop:direction}, which is all the direction of the hazard needs, is identified whatever $\E Z$ is.
\end{remark}

\paragraph{Proof of \cref{lem:a0}.} $\E\,y_i(y_i-1)=\sum_{a\neq b}(p_{ia}p_{ib}+C_{ab})=(n_i\bar p_i)^2-\sum_ap_{ia}^2+n_i(n_i-1)\bar C_i$ and $\sum_ap_{ia}^2=n_i\bar p_i^2+n_i\Var_{\mathrm{pop}}(p_{ia})$; dividing by $n_i(n_i-1)$ gives $\bar p_i^2+\bar c_i$. Step (i) depends only on $\E\hat p_i=\bar p_i$. Step (ii) solves $R(\delta)=\sum_iw_i\E\hmtwo_{2,i}=\sum_iw_i\bar p_i^2+\Delta_w$ with $R(\delta)=\sum_iw_i\expit(\hat\alpha+\hat\gamma x_i+\delta)$ strictly increasing.

\paragraph{Proof of \cref{lem:cross}.} For runs $a,b$ in different batches, $\E[Y_{ia}Y_{ib}\mid p_i,\text{batches}]=\E[Y_{ia}\mid p_i]\,\E[Y_{ib}\mid p_i]=p_i^2$. Under factorisation the density of $Z$ given $t$ and selection is proportional to $f_Z(z)\psi_z(z)$, which does not depend on $t$, so the selected tasks follow $\cM$ with the same $H_0$ and \cref{prop:shift} applies. For non-necessity, take $\theta=1$ and $\psi(z,x)=\tfrac12+\tfrac14\sin(2\pi x)(1-e^{-z})$: it does not factorise, but it is invariant under $x\mapsto x+\theta$, and so is the law of $Z$ given $x$ and selection; with $H_0(x+1)=2H_0(x)$ this gives (T). The same argument, with a retention probability $\chi(z,t)$ in place of $\psi$ and $0<\int f_Z\chi\,dz<\infty$, covers outcome-based filtering.

\begin{lemma}[Composition]\label[lemma]{lem:comp}
Let sources $j$ have shares $\xi_j(x)$ with $\sum_j\xi_j\equiv1$ under the weighted task measure, within-source curves $S_j$ with $m_{2,j}(x)=S_j(x+\theta_j)$, and pooled curves $S=\sum_j\xi_jS_j$, $m_2=\sum_j\xi_jm_{2,j}$. Then for any reference shift $\theta$
$m_2(x)-S(x+\theta)=\sum_j[\xi_j(x)-\xi_j(x+\theta)]S_j(x+\theta)+\sum_j\xi_j(x)[S_j(x+\theta_j)-S_j(x+\theta)]$.
(T) is preserved if $\theta_j\equiv\theta$ and either the shares do not depend on $x$ or the source curves coincide; otherwise its violation can have either sign.
\end{lemma}

\begin{remark}[Shared frailty, and the power-law case; for \cref{sec:diag}]\label[remark]{rem:frailty}
Suppose $-\log P_i(s\mid c)=Z_iH(s\mid c)$ with $Z_i$ shared across lengths. If $g(c)=H(c\mid c)$ is non-decreasing on $\mathcal C$, then $Z_ig(s)$, extended between design lengths, is a twin. For design lengths $a<b$ with positive terms,
\[
\log g(b)-\log g(a)=\underbrace{\log H(b\mid b)-\log H(a\mid b)}_{\text{within the longer task}}+\underbrace{\log H(a\mid b)-\log H(a\mid a)}_{\text{length effect at the common step } a},
\]
an exact split of the diagonal's slope into a within-run change and a length effect. In the continuum, and if $H$ has a local $C^1$ extension, this becomes $d\log g/d\log c=k_w(c)+\eta(c)$, where $\eta$ is the derivative in $\log c$ at fixed step. The power law $H(s\mid c)=(s/a)^k(c/a)^\beta$ gives $k_w=k$ and $\eta=\beta$. Its diagonal is the Weibull of shape $k+\beta$ ($k+\beta>0$), which is the end-of-task equivalence of \citet[App.~B.3]{anon2026pooling}. Their tasks differ in rated length, and the two-dimensional $H$ here is our extension: their data do not observe one task at several lengths. A related observational-equivalence idea in duration analysis is that no anticipation cannot be tested from observational data \citep{abbring2003nonparametric}. The experiment and the restrictions there differ from ours, and neither result implies the other. \citet{bhagia2023} identifies duration dependence from two notice lengths by assuming that the notice length affects only the first period. Here the length effect is the object to be tested.
\end{remark}

\paragraph{Pooled ratios and scale type.} For \cref{prop:scale}: with a gamma frailty of mean and variance one, the pooled cumulative hazard at step $j$ is $\log\{1+H(j\mid c)\}$, so under a constant factor $a$ the pooled ratio is $\log(1+aH)/\log(1+H)$, which varies with $H$.

\paragraph{A0 does not give the paired test.} Take $n=2$ and let the counts of the two arms of one instance be $(0,1)$, $(1,2)$, $(2,0)$ or $(1,1)$, each with probability $1/4$, the order of runs within an arm being a uniform permutation. Each arm is $\mathrm{Bin}(2,1/2)$ and satisfies A0, but the difference takes the values $-1,-1,2,0$ and is not symmetric about zero. \Cref{thm:test}(3)--(4) therefore assume A0$^{+}$.

\section{Preregistrations and deviations}\label{app:prereg}

Each experiment was specified in a dated preregistration document (in Chinese), and its design, driver and primary analysis code were frozen, with SHA-256 hashes recorded, before any of its outcomes existed; one descriptive analysis was added to S1 while it was running, before any of its outcomes had been read (below). The documents were not deposited with an external registry, so the order of events rests on our dated records; the hashes identify the files but do not certify when they were written. The documents, frozen copies of the code, hashes, raw outputs and analysis logs are in the code release. A Tower of Hanoi pilot and a design calibration preceded Pilot~A and informed its choice of task and lengths. In the pilot (50 instances, 8 runs each at 44 moves), instances differed only by a relabelling of discs, success was almost unrelated across runs of the same instance (intraclass correlation $0.010$), and step 32, where the rule was misapplied, was the most common first-error position (68 of 148 failures at temperature 1). A task with so little heterogeneity across instances cannot show the difference between cross-sectional and within-run shapes, so we switched to nested linear maps. The pilot is not reported in \cref{sec:experiments}; its preregistration and results are in the release.

\paragraph{Pilot A} (frozen 2026-11-27). Written after the Hanoi pilot and the design calibration had been seen, and before any Pilot~A outcome. The task (nested linear maps), the lengths $\{24,48\}$ and temperature $1$ were chosen in view of those results, and the second temperature was dropped to control cost. Primary checks: (W) at $c=48$, $t=24$ and (P1) at $j=24$, Holm-adjusted. The comparison of cross-sectional and within-run shapes with the shift estimator of \cref{sec:repeated} was preregistered as descriptive. Its text named a per-instance Weibull fit to first-error times as the within-run side; the frozen code, and what we report, use the within-run shift estimator instead. The frailty-NPMLE comparison on Pilot~A was run after its data had been seen, and it motivated the wider grid for $k$ used later.

\paragraph{S1} (frozen 2026-11-28, before any S1 outcome). Primary checks: a replication gate on (W) and (P1) at temperature $1$; (W), (P1) and the requested-length contrast under recommended decoding, Holm-adjusted; and a dose-response check by an intersection--union test of the increments $24\to36$ and $36\to48$, declared in advance not to be a gate. Deviations from the earlier plan, all made before any S1 data:
\begin{itemize}\setlength{\itemsep}{0pt}
\item The dose-response check was moved from $j=12$ to $j=24$, because simulated power at $j=12$ was about zero.
\item Recommended decoding was restricted to one arm, and the input-length arm to that decoding, to control cost.
\item One scenario of the synthetic check of the analysis code was mis-specified and corrected; the analysis code itself was unchanged.
\end{itemize}
The preregistration had planned to reuse S1's instances in S2; S2 used new instances instead, which is why the three sets are independent. One revision was made while S1 was running, before any of its outcomes had been read (only token counts were inspected). It added the frailty-NPMLE comparison of \cref{sec:experiments} as a descriptive analysis, with its code frozen at that point, and widened the grid for $k$ of \citet{anon2026pooling} from $[0.3,4]$ to $[0.3,20]$.

\paragraph{S2} (frozen 2026-11-28, before any S2 outcome, and re-frozen the same day before any data after timestamps in code comments were corrected). The design was chosen after S1's results and a 160-request calibration, which also fixed the lengths of the independent arm by a rule written in advance. The arms with the added sentence were added because S1's input-length contrast had contained that sentence. Primary check: whether the length effect is larger on the dependent chain than on independent expressions, judged on both the probability and the logit scale. The secondary family (Holm, four tests) comprised re-asking the second half as a new task, the requested-length contrast, the list-length contrast and the added sentence. The arm with lists of length $48$ and the added sentence was included so that all three contrasts are estimated within one batch. The decision table for the primary check did not anticipate the outcome observed, a rejection on the probability scale with the logit-scale interval entirely below zero. This check was preregistered as the one distinguishing our setting from \citet{li2026fragility}. Read on the probability scale alone it would pass; on the stricter reading it does not. \Cref{sec:experiments} reports the numbers, calls the comparison inconclusive, and gives a reason found after seeing the data, a confound with input format. A synthetic scenario of the decision table had its expected sign written wrongly in a comment; this was corrected before any data.

\paragraph{G} (frozen 2026-11-29, before any output of the model, including its calibration). Written after all results on Qwen3-8B had been seen. Model gpt-oss-20b at a fixed revision, its own chat format with medium reasoning effort, temperature $1$ and top-$p$ $1$ as its developers recommend, the same prompts and the same scoring, with the model's final answer channel in place of the text after the reasoning. A calibration on 20 separate instances applied three rules written in advance: a parsing gate; a length rule, which moved the lengths to $48$ and $96$ because at least 95\% of calibration runs at length $48$ had their first $24$ steps right; and a budget rule, which chose the four-arm design. Primary checks: (W) and (P1), Holm-adjusted, as in Pilot~A. Secondary family (Holm, three tests): list length, the added sentence and requested length, as in S2. There were no deviations from the registered rules and analyses; the two checks on unparsable runs reported in \cref{sec:experiments} are post hoc.

\paragraph{Analyses not preregistered.} The frailty-NPMLE comparison on Pilot~A, the null calibration of that estimator, the unpaired re-splitting comparison, the instruction-corrected requested-length effect, and the two checks on unparsable runs in G were all written after the data had been seen. So was a rule for classifying first errors: the S1 preregistration had listed index slips among its exploratory failure categories, but the matching rule and the distant-offset comparison were written after the main analysis. A first wrong value reported at step $k\ge1$ counts as consistent with an index slip if it can be reproduced from the correct previous value by taking one coefficient, or both with the same offset, from a list entry one or two positions away. Under recommended decoding in S1, 11 of the 14 such first errors on steps 19--24 are consistent with an index slip at length 24, and 74 of the 100 at length 48. %
Pooling the chain's lengths and steps under recommended decoding, 727 of the 1,066 first wrong values with at least six list positions on each side match a near offset (one or two positions), and 402 of the same 1,066 match a distant offset (five or six); the two are not exclusive. %
Both lengths show the pattern, and the classification does not distinguish harder lookup in a longer list from a longer reasoning trace. Where they appear in the text, they are labelled as post hoc.

\section{Pipeline details and per-model results}\label{app:pipeline}

\paragraph{Status gate and marks.} The gate rejects, in order: no information, rank deficiency, (quasi-)separation, a second moment at the boundary, non-convergence or numerically unreliable fits, a non-negative slope, and sensitivity to the ridge ($|\htheta(10^{-6})-\htheta(10^{-9})|>0.01\max(1,|\htheta|)$). Every bootstrap draw is gated; an interval with more than 5\% failed draws is marked and excluded from counts. In TH1.0 this marks all ten estimable cells of GPT-2 (5 task families, all 66 tasks from SWAA).

\paragraph{Specifications.} Base; no SWAA; HCAST only; RE-Bench only (never estimable); $t\ge2$, $t\ge8$ and $t\le240$ minutes; fatal runs kept; equal weights; METR's run weights; tasks with one task version; no cloned tasks; baseline-timed tasks only; estimated times only; and two outcome-based filters reported for comparison only.

\paragraph{Calibration.} We simulate from $\cM$ on observed designs with $\beta=0.8$ and $k\in\{1,2\}$. In the check of outcome-based filtering (all models of a release with a shared task frailty, 30 simulations per cell, no bootstrap), each simulation records the median of the per-model estimates; the mean of this median minus the true $\theta$ is between $-0.021$ and $+0.010$ without filtering and between $+0.149$ and $+0.322$ with either filter. In the coverage check (40 simulations per cell, 200 bootstrap draws each, on the designs of GPT-5.2 in TH1.1 and GPT-5.1-Codex-Max in TH1.0) the mean error is between $+0.035$ and $+0.062$, with simulation standard deviations near $0.2$, and percentile intervals cover the truth in $0.88$--$0.95$ of simulations. These are illustrations on a few templates, not a calibrated coverage guarantee.

\begin{table}[ht]
\centering
\small
\caption{Frontier models: point estimate and 95\% interval for $1/k$ (endpoints rounded outward), and the point estimate of $1/\beta$ from the first step; within $\cM$, $1/\beta<1$ already implies $k>1$. ``--'': too few tasks with cross-batch pairs. TH1.1 model names omit the suffix ``(Inspect)'' carried by some of them in the data. The all-pairs column uses all tasks with $n_i\ge2$, the cross-batch column only tasks with runs in at least two batches.}\label{tab:frontier}
\begin{tabular}{@{}llrrll@{}}
\toprule
Release & Model & $\hat h$ (min) & $1/\hat\beta$ & all pairs (base) & cross-batch pairs \\
\midrule
TH1.0 & o4-mini & 100 & 1.083 & 0.392 [0.214, 0.683] & 0.375 [0.198, 0.762] \\
 & o3 & 107 & 0.937 & 0.392 [0.184, 0.708] & 0.388 [0.172, 0.763] \\
 & GPT-5 & 121 & 0.901 & 0.270 [0.117, 0.478] & 0.304 [0.124, 0.535] \\
 & Claude Opus 4.5 & 132 & 1.004 & 0.445 [0.205, 0.799] & 0.433 [0.162, 0.788] \\
 & GPT-5.1-Codex-Max & 195 & 1.133 & 0.241 [0.038, 0.573] & 0.243 [0.030, 0.608] \\
\midrule
TH1.1 & Claude 4 Opus & 100 & 1.148 & 0.742 [0.512, 1.071] & 0.673 [0.368, 1.189] \\
 & Claude 4.1 Opus & 100 & 1.049 & 0.566 [0.363, 0.829] & -- \\
 & o3 & 188 & 1.107 & 0.515 [0.287, 0.846] & 0.462 [0.214, 0.880] \\
 & Gemini 3 Pro & 218 & 0.993 & 0.487 [0.269, 0.800] & -- \\
 & GPT-5 & 241 & 1.259 & 0.462 [0.265, 0.761] & 0.446 [0.226, 0.818] \\
 & Claude Opus 4.5 & 294 & 1.249 & 0.357 [0.210, 0.555] & -- \\
 & GPT-5.1-Codex-Max & 336 & 1.156 & 0.517 [0.274, 0.849] & -- \\
 & GPT-5.3-Codex & 465 & 1.441 & 0.538 [0.304, 0.919] & 0.517 [0.272, 0.966] \\
 & Claude Opus 4.6 & 664 & 1.649 & 0.533 [0.292, 0.927] & 0.570 [0.275, 1.052] \\
 & GPT-5.2 & 1029 & 1.500 & 0.446 [0.199, 0.988] & -- \\
\bottomrule
\end{tabular}
\end{table}

\section{Provenance of numbers}\label{app:prov}

Numbers in the paper are of four kinds. (i) Empirical summaries, the mechanism examples of \cref{tab:mech}, the illustration in \cref{sec:cross} and the cross-sectional slopes $\hat\beta$ are printed by \texttt{s3\_paper\_numbers.py}, which reads the JSON ledgers of the analysis scripts (\texttt{metr\_s1\_robustness.py}, \texttt{metr\_s1\_a0\_pairs.py}, \texttt{metr\_s1\_simex.py}; SHA-256 hashes in its output) and quotes the archived outputs of the deterministic checks; its counts for \cref{tab:counts} are compared line by line with an independent reconciliation script. The examples in \cref{rem:three} are checked by \texttt{s3\_review\_checks.py}. (ii) Sums and counts derived from these, such as $11=8+3$ rows or eight of twelve specifications, are stated as such. (iii) Protocol constants (95\% intervals, the 5\% failed-draw threshold, ridge values, batch gap, minimum number of tasks) come from the analysis code. (iv) Numbers quoted from other work, such as the Weibull shapes in \cref{sec:intro}, come from the cited sources. For these numbers a table mapping each to its source field and rounding rule accompanies the code; interval endpoints are rounded outward and point estimates to nearest. Numbers in \cref{sec:experiments} and \cref{app:prereg} are different: they are printed to three decimals by the frozen analysis scripts and by provenance scripts that quote those outputs, and are copied from the printed lines, each with a tag; a map from tags to sentences is in the code release. Bounds stated as `at least' or `at most' are computed exactly from the run-level data and rounded outward.

\end{document}